\documentclass[floatfix,superscriptaddress,preprint]{revtex4-1}
\usepackage[english]{babel}
\usepackage{graphicx}
\usepackage{float}
\usepackage{dcolumn}
\usepackage{bm}
\usepackage{float}
\usepackage{caption}
\usepackage{subcaption}
\usepackage{gensymb}
\usepackage{epsf}
\usepackage{xcolor}
\newcommand\twoFigureSize{0.47\linewidth}

\newcommand\SCALE{0.59}

\begin{document}
\title {Coexistence of attractors in a coupled nonlinear delayed system modelling El Ni\~{n}o Southern Oscillations}
\author{Chandrakala Meena\footnote{email : chandrakala@iisermohali.ac.in}} 
\affiliation{Indian Institute of Science Education and Research (IISER) Mohali, Knowledge City, SAS Nagar, Sector 81, Manauli PO 140306, Punjab, India}
\author{Elena Surovyatkina\footnote{elena.surovyatkina@gmail.com}}
\affiliation{Potsdam Institute for Climate Impact Research, Telegrafenberg A31 14473 Potsdam, Germany}
\affiliation{Space Research Institute, Russian Academy of Sciences, Profsoyuznaya 84/32, GSP-7, Moscow 117997, Russia}
\author{Sudeshna Sinha\footnote{email : sudeshna@iisermohali.ac.in}}
\affiliation{Indian Institute of Science Education and Research (IISER) Mohali, Knowledge City, SAS Nagar, Sector 81, Manauli PO 140306, Punjab, India}

\begin{abstract}
	We study the dynamics of the sea surface temperature (SST) anomaly using a model of the temporal patterns of two sub-regions, mimicking behaviour similar to El Ni\~{n}o Southern Oscillations (ENSO). Specifically, we present the existence, stability, and basins of attraction of the solutions arising in the model system in the space of these parameters: self delay, delay and inter-region coupling strengths. The emergence or suppression of oscillations in our models is a dynamical feature of utmost relevance, as it signals the presence or absence of ENSO-like oscillations. In contrast to the well-known low order model of ENSO, where the influence of the neighbouring regions on the region of interest is modelled as external noise, we consider neighbouring regions as a coupled deterministic dynamical systems. Different parameters yield a rich variety of dynamical patterns in our model, ranging from steady states and homogeneous oscillations to irregular oscillations and coexistence of oscillatory attractors, without explicit inclusion of noise. Interestingly, if we take the self-delay coupling strengths of the two sub-regions to be such that the temperature of one region goes to a fixed point regime when uncoupled, while the other system is in the oscillatory regime, then on coupling both systems show oscillations. This implies that oscillations may arise in certain sub-regions through coupling to neighbouring regions. Namely, a sub-region with very low delay, which would naturally go to a steady state when uncoupled, yields oscillations when coupled to another sub-region with high enough delay. 
	
	We explicitly obtain the basins of attraction for the different steady states and oscillatory states in the model. Our results might be helpful for forecasting of El Ni\~{n}o (or La Ni\~{n}a) progress, as it indicates the combination of initial SST anomalies in the sub-regions that can result in a El Ni\~{n}o/La Ni\~{n}a episodes. In particular, the result suggests using an interval as a criterion to estimate the El-Nino or La-Nino progress instead of the currently used the single value criterion.

	\end{abstract}
	
	\maketitle
	\section{Introduction}
	
	 El Ni\~{n}o is an ocean-atmospheric event, occurring at intervals of two to seven years. It has attracted much popular interest as it has global impact that ranges from environment to economics. 
	   The El Ni\~{n}o Southern Oscillations (ENSO) typically signifies an irregular cycle of coupled ocean temperature and atmospheric pressure oscillations across the equatorial Pacific region, with one phase involving large scale warm events marked by dramatic change in sea surface temperature (SST)  \cite{Andrews,Encyclopedia,Holton_Hakim,review}.
	  
	  In normal years, SST of the western Pacific Ocean is high and pressure is low compared to the eastern Pacific Ocean. Due to high SST in the western region, evaporation increases and high rainfall occurs there. Less rainfall occurs in the east due to cold SST and high pressure levels. A pressure gradient in the east and west pacific ocean induces circulations of trade winds. These circulating trade winds in turn affect the depth of the thermocline gradient. In normal conditions, the thermocline is deeper in the western Pacific region and shallower in the eastern region. However when the El Ni\~{n}o becomes very strong, the circulation of trade winds changes its direction. As a result the thermocline depth becomes almost the same in both east and west Pacific Ocean. In contrast to El Ni\~{n}o, La Ni\~{n}a is the cold phase of ENSO, with the cycle of hot and cold phases having an average periodicity of approximately 3.7 years. 
	
	The first modern mechanism underlying ENSO was proposed by Bjerknes. He hypothesized that positive feedback between the atmosphere and the equatorial eastern Pacific ocean leads to the El Ni\~{n}o effect \cite{bjerknes}. Now positive ocean-atmosphere feedback is responsible for the growth of internal instabilities, that can produce very large SST anomalies in the eastern tropical Pacific region. To keep the instability in the SST anomalies bounded, negative feedback is necessary. Therefore to gain understanding of the positive-negative feedback mechanisms underlying the emergence of ENSO, several low order models (LOM) have been introduced in the past decades. For instance, one of the earliest efforts to obtain ENSO-like oscillations was proposed by Zebiak and Cane \cite{cane}, and the effect of the ocean and atmosphere on each other was central to their model. Based on the coupled model of Zebiak and Cane, the recharge oscillator model was proposed by Jin, based on the recharge and discharge process of warm water over tropical Pacific ocean \cite{Jin_1997a,Jin_1997b}. Subsequently, consistent with the observations of ENSO, the western Pacific oscillator model \cite{Wang_1997,Wang_1999} was proposed, where the role of the western Pacific in ENSO was emphasized. Other attempts include that by Picaut, who introduced an advective-reflective oscillator, which includes a positive feedback of zonal currents that advect from the western Pacific warm pool toward the east during El Niño \cite{Picaut_1997}. The main motivation of such simple models is to gain understanding of the underlying mechanisms of ENSO, through basic models involving a small number of variables, that are capable of provide qualitative description of the complex phenomenon of ENSO \cite{ Kirtman_1998,Ghil,Seon_2012,tziperman}.
	 	 	 
	 	One important class of models attempting to understand the behaviour of ENSO is the deterministic low order  {\em delayed action oscillator model} \cite{delay,battisti_hirst}. This model will be the focus of this work. Delayed negative feedback models provide a very good, yet simple, representation of the basic mechanism of ENSO-like oscillations. An important feature of this class of models is the inclusion of a delayed feedback which incorporates oceanic wave transit effects, namely the effect of trapped ocean waves propagating in a basin with closed boundaries. Specifically, the delayed-action oscillator model has three terms, and is a first  order nonlinear delay differential equation for the temperature anomaly $T$, i.e. the deviation from a suitably long term average temperature, given by:
	 
	 	\begin{equation}
	 		\frac{dT}{dt}=k T- b T^3- A T(t-\Delta)
	 		\label{basic}
	 	\end{equation}
	 	Here the coupling constants are $k$, $b$ and $A$, with $\Delta$ being the delay. The first term represents a positive feedback in the ocean-atmosphere system, working through advective processes giving rise to temperature perturbations that result in atmospheric heating. The heating in turn leads to surface winds driving the ocean currents which then enhance the anomalous values of $T$. The second term is a damping term, due to advective and moist processes, that limits the temperatures from growing without bound. The delay term arises from considerations of equatorially trapped ocean waves propagating across the Pacific and interacting back after a time delay, determined by the width of the Pacific basin and wave velocities. The strength of this interaction, relative to the nondelayed feedback is given by $A$.
	 
	 	We will consider the dimensionless form of this equation \cite{AmJPhys}:
	  	\begin{equation}
	 		\frac{dT}{dt}=T-T^3-\alpha T(t-\delta)
	 		\label{main1}
	 	\end{equation}
	 	
	 	where time in Eqn.~1 has been scaled by $k$, temperature by $\sqrt{b/k}$. The dimensionless constants $\alpha = A/k$ and $\delta = k\Delta$ \cite{AmJPhys}. This  model allows multiple steady states and when these fixed points become unstable, self-sustained oscillations emerge. Thus this class of models provide a simple explanation of ENSO, and provides insights on the key features that allow the emergence of oscillatory behavior. 
	 
	 %	\section{Coupled Delayed-Oscillator Model}
	 	The delayed-oscillator model given by Eqns~1-2 above consider a single region with strong atmospheric-ocean coupling, namely some typical representative region in the Pacific Ocean. This approach was taken further in Ref.~\cite{csf_ms} where two sub-regions of the Pacific was incorporated in the model. Specifically, these models mimicked the coupling of regions along the equator, where one expects varying self-delay coupling strengths in the sub-regions, as well as varying (possibly strong) delay times \cite{AmJPhys}. We will first describe this coupled model in detail, and then review the patterns emerging in this class of models, as obtained in Ref.~\cite{csf_ms}. 
	 	
	 	Consider two coupled sub-regions, given by following dimensionless delay differential equations, as introduced in \cite{AmJPhys,csf_ms}:	
	 	\begin{eqnarray} 
	      \label{main}
	 	\frac{dT_{1}}{dt}= T_{1}-T_{1}^3-\alpha_{1} T_{1}(t-\delta_{1})+\gamma T_{2}\\ \nonumber
	 	\frac{dT_{2}}{dt}= T_{2}-T_{2}^3-\alpha_{2} T_{2}(t-\delta_{2})+\gamma T_{1}
	 	\end{eqnarray} 
	 	
	 	Here $T_{i}$, $\delta_{i}$ and $\alpha_{i}$ with $i=1,2$ are the scaled temperature anomaly, self-delay, strength of the self-delay of each sub-region, and $\gamma$ is the inter-region coupling strength between the two regions. The form of the coupling term models the situation where if one region is cooler than the other, then the flow of energy across their common boundary will result in heating one sub-region and cooling the other. We give below the salient dynamical features arising in this system.

    \section{Dynamics of coupled identical sub-regions}
\begin{figure}[H]
			\centering 
			\includegraphics[scale=\SCALE]{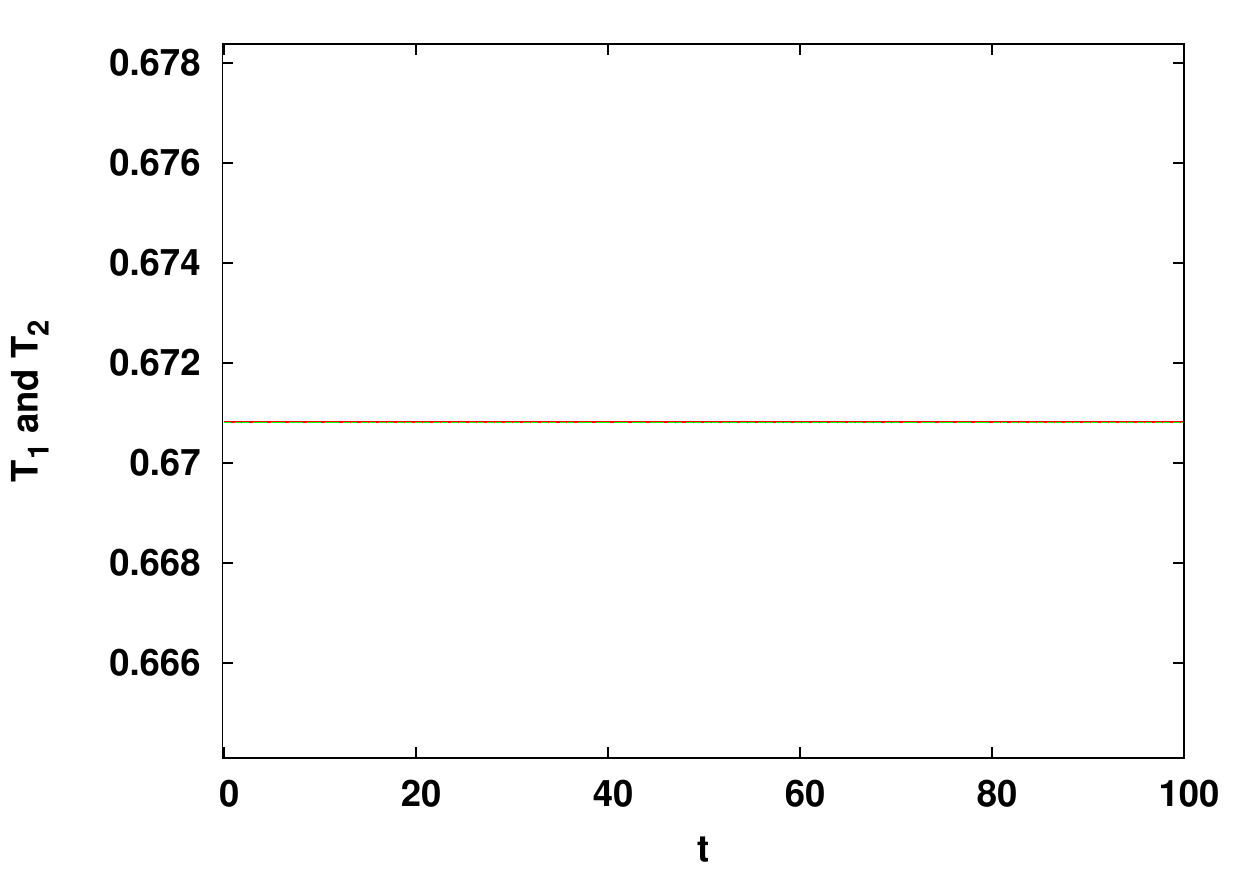}
			\includegraphics[scale=\SCALE]{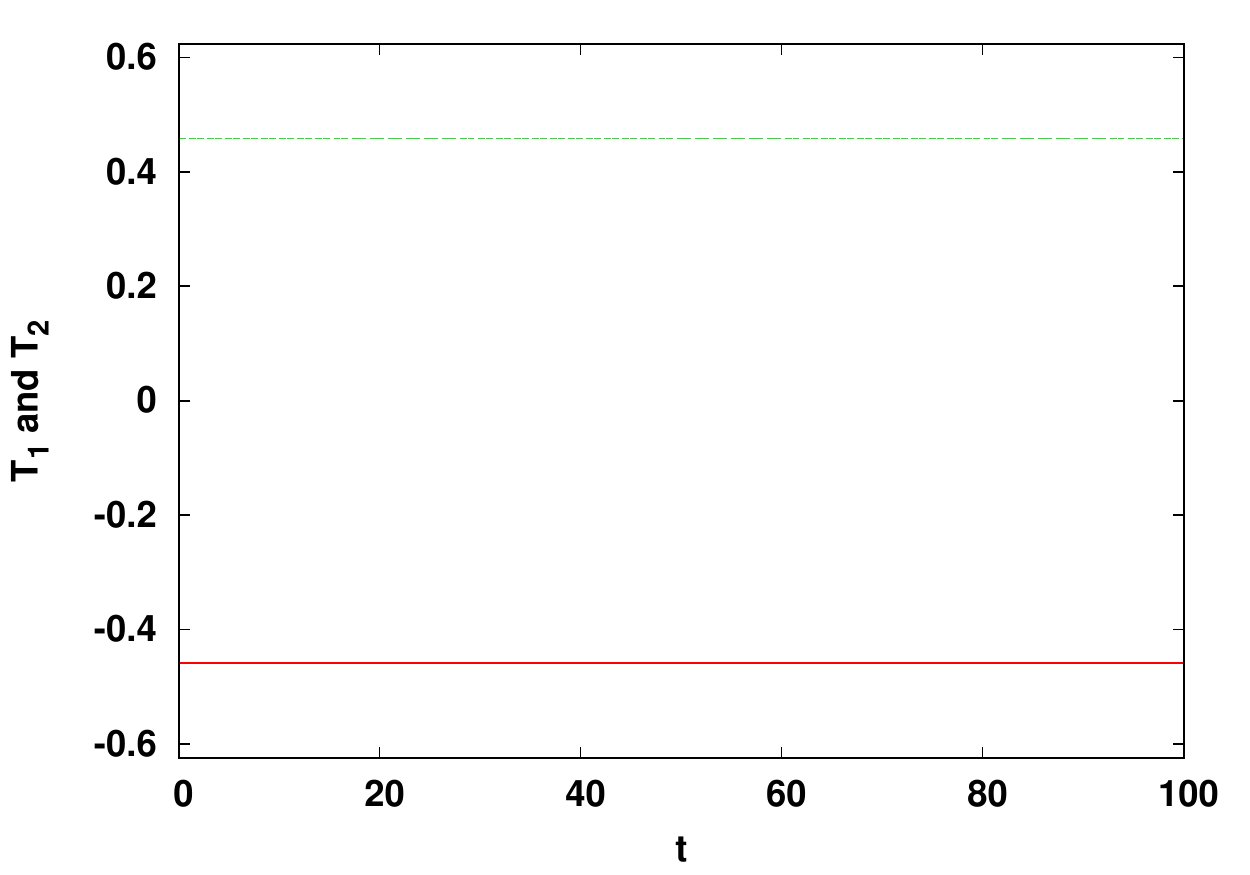}
			\caption{Temporal evolution of the temperature anomalies of the two sub-regions $T_1$ (in red) and $T_2$ (in green) with $\alpha_{1} = \alpha_{2} = 0.75$, $\delta_{1}=\delta_{2}=1$, and inter-region coupling (right) $\gamma=0.2$  and (left) $\gamma=0.05$, showing amplitude death and oscillator death behavior respectively \cite{csf_ms}.}
			\label{identical1}
	\end{figure}

    %\vspace{-1cm}
    First we consider the case of identical sub-regions, i.e. $\alpha_{1}=\alpha_2$ and $\delta_{1}=\delta_2$. This arises when the two regions are geographically close-by, and the distance from the western boundary is approximately same, with the same losses and reflection properties for both regions and similar transient time taken by the oceanic waves. Four distinct types of behaviour emerge in this case:
	
	(i) Amplitude Death (AD) : here both regions go to a single steady state  \cite{AD_OD}. See left panel of Fig. \ref{identical1}  for a representative example.
	
	(ii) Oscillation Death (OD): here the sub-regions go to different steady states \cite{AD_OD}. See right panel of Fig. \ref{identical1}  for a representative example.
	
\begin{figure}[H]
			\centering 
			\includegraphics[scale=\SCALE]{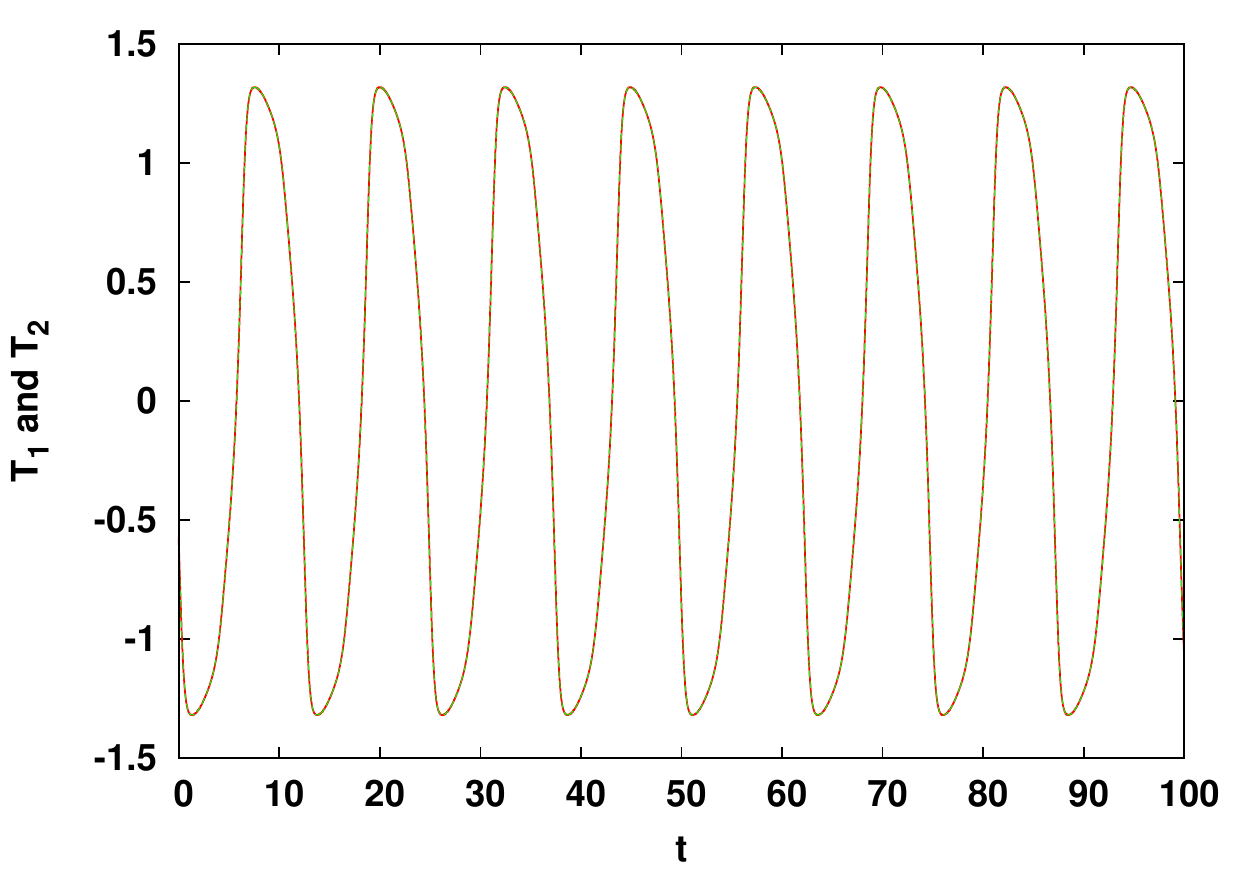}
			\includegraphics[scale=\SCALE]{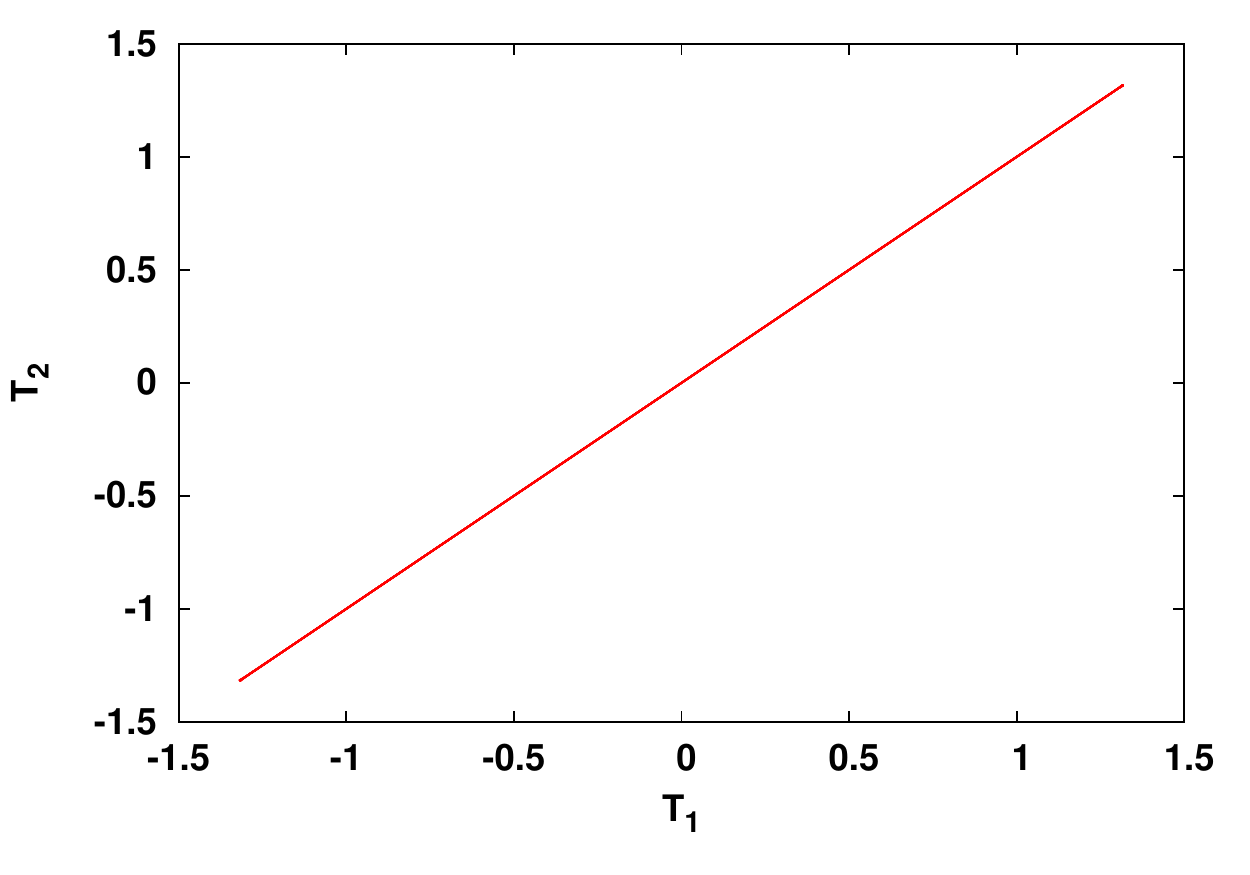}
			\caption{Temporal evolution of the temperature anomalies of the two sub-regions $T_1$ (in red) and $T_2$ (in green) in the left panel, and the corresponding phase portrait in the $T_1-T_2$ plane in the right panel, for $\alpha_{1} = \alpha_{2} = 0.75$, $\delta_{1}=\delta_{2}=4$ and $\gamma=0.1$ in Eqn. \ref{main}  \cite{csf_ms}.}
	       \label{identical2}		
	\end{figure}

	(iii) Homogeneous oscillations : here the regions oscillate synchronously and there is no phase or amplitude difference between the oscillations. See Fig. \ref{identical2} for a representative example.
	
	(iv) Heterogeneous oscillations : here the oscillatory patterns are complex, and the oscillations in the two sub-regions differ in either phase or amplitude, or both. Further, the oscillations may be irregular for certain parameters. See Figs. \ref{delta_alpha_1}-\ref{patterns}  for representative examples.
	
	It was evident from our extensive numerical simulations that oscillations emerge as the delay $\delta$ and strength of self-delay coupling $\alpha$ increases, and as inter-region coupling strength $\gamma$ decreases. Importantly, as compared to a single region model, oscillations emerged for larger values of delay in the two coupled sub-regions model. {\em This implies that coupling of sub-regions yields smaller parameter regions giving rise to ENSO-like oscillations.}
	
	Note that in our model system we have observed that the number of attractors and their basins of attraction depend upon the values of  parameters. For instance, when $\alpha_{1}=\alpha_{2}=0.75, \delta=1$, we find four steady states for $\gamma=0.1$ and two steady states for $\gamma\ge0.2$. The value of the fixed points depend on the values of the inter-region coupling strength $\gamma$. For the typical case of $\alpha_1 \ne \alpha_2$, each region has two fixed points and two oscillator states, with the attractors being different in the two regions. Generically, in such cases there is a complex co-existence of attractors.  We will investigate this aspect in greater detail in the section below.

	As the strength of the inter-region coupling $\gamma$ increases, co-existence of AD and OD decreases. Further, the region of amplitude death increases, implying that the ENSO is less likely when two sub-regions are strongly coupled. We also observe that as delay $\delta$ increases, co-existence of AD and OD decreases, and the parameter region supporting oscillatory behaviour increases. For instance, when $\delta=2$ oscillations emerge for self-delay coupling strength $\alpha \geq 0.65$, while for $\delta=4$ oscillations emerge in the systems with $\alpha \geq 0.48$. So longer delays, namely longer oceanic wave transit times, favour ENSO-like oscillations.
	\section{Dynamics of coupled non-identical sub-regions}
\begin{figure}[h]
			\centering 
			\includegraphics[scale=\SCALE]{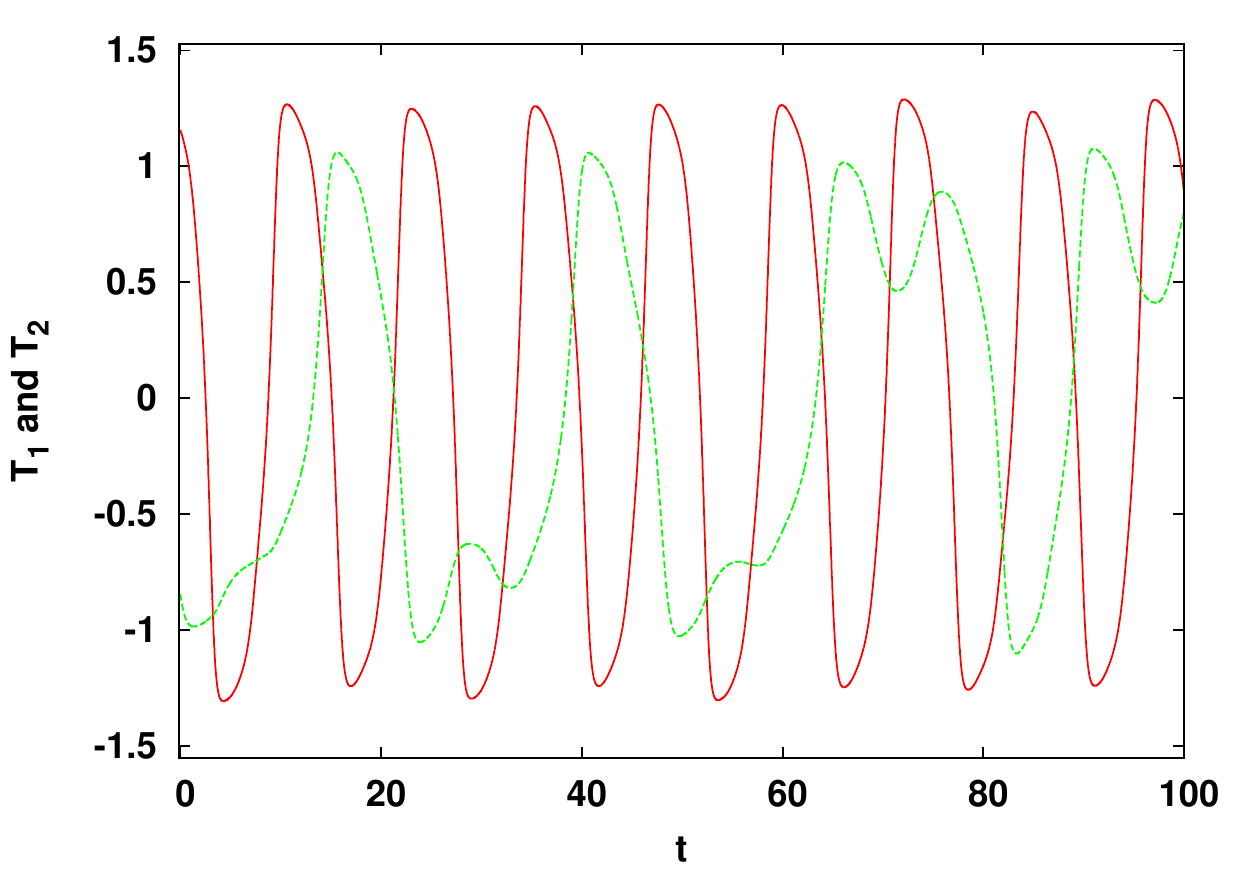}
			\includegraphics[scale=\SCALE]{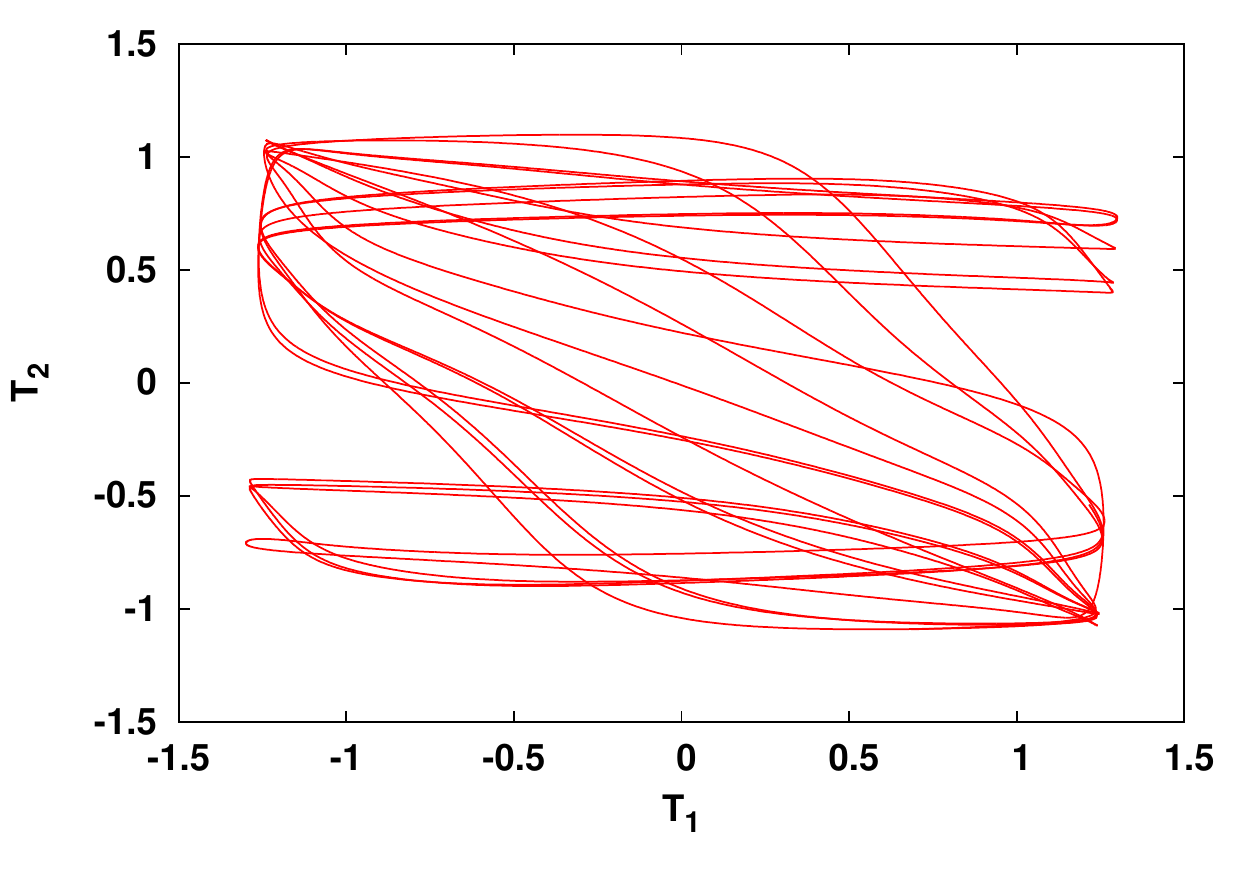}
			
			(a) 
			\vspace{1cm}
			
			\includegraphics[scale=\SCALE]{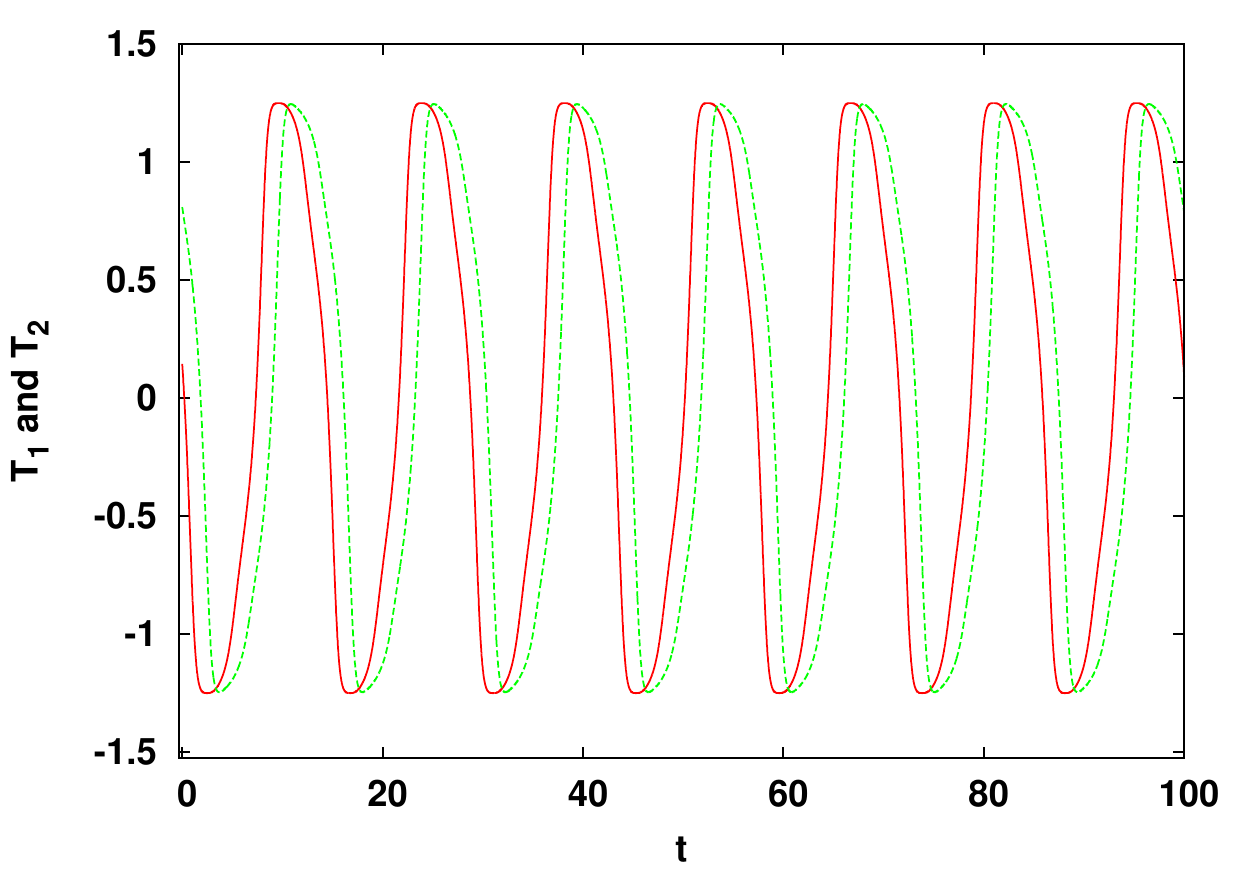} 
			\includegraphics[scale=\SCALE]{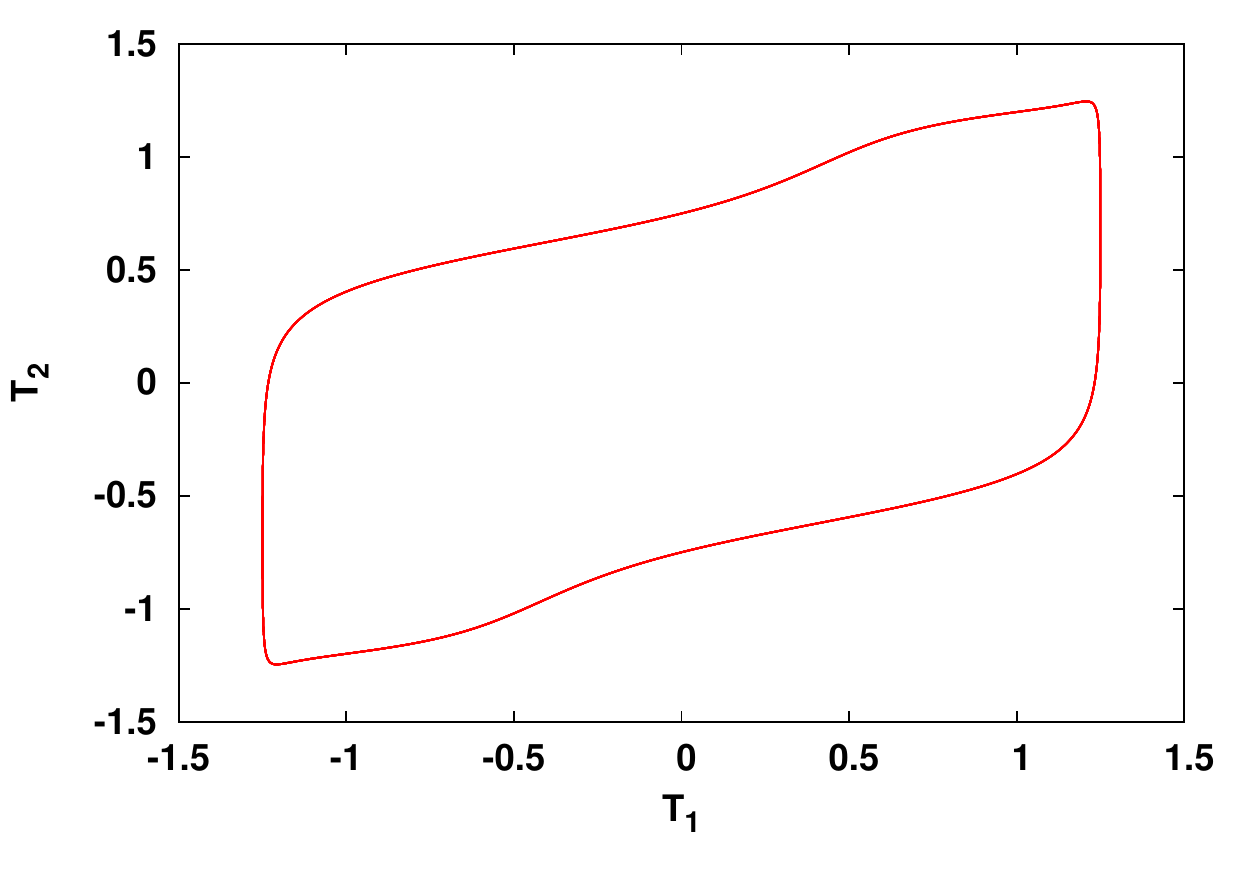}
			
			(b)
	
			\caption{Temporal evolution of the temperature anomalies of the two sub-regions $T_1$ (in red) and $T_2$ (in green), in the left panels, and the corresponding phase portraits in the $T_1-T_2$ plane in the right panels, for a system with $\alpha_{1} = 0.75, \alpha_{2} = 0.5$, coupling delay $\delta =4$ and inter-region coupling strength  $\gamma$ equal to (a) $0.1$ and (b)  $0.2$, in Eqn.\ref{main}  \cite{csf_ms}.}	\label{delta_alpha_1}
		\end{figure}

			Now we will consider the case of non-identical sub-regions, i.e. $\alpha_{1} \ne \alpha_2$ and $\delta_{1} \ne \delta_2$, relevant to the case where the distance from the western boundary is different for the sub-regions and therefore the transient times taken by the oceanic waves are different in the sub-regions.  Fig.~\ref{delta_alpha_1} shows the typical dynamics emerging under varying differences in the two sub-regions $\Delta \alpha = \alpha_1 - \alpha_2$.
			
\begin{figure}[H]
			\centering 
			\includegraphics[scale=\SCALE]{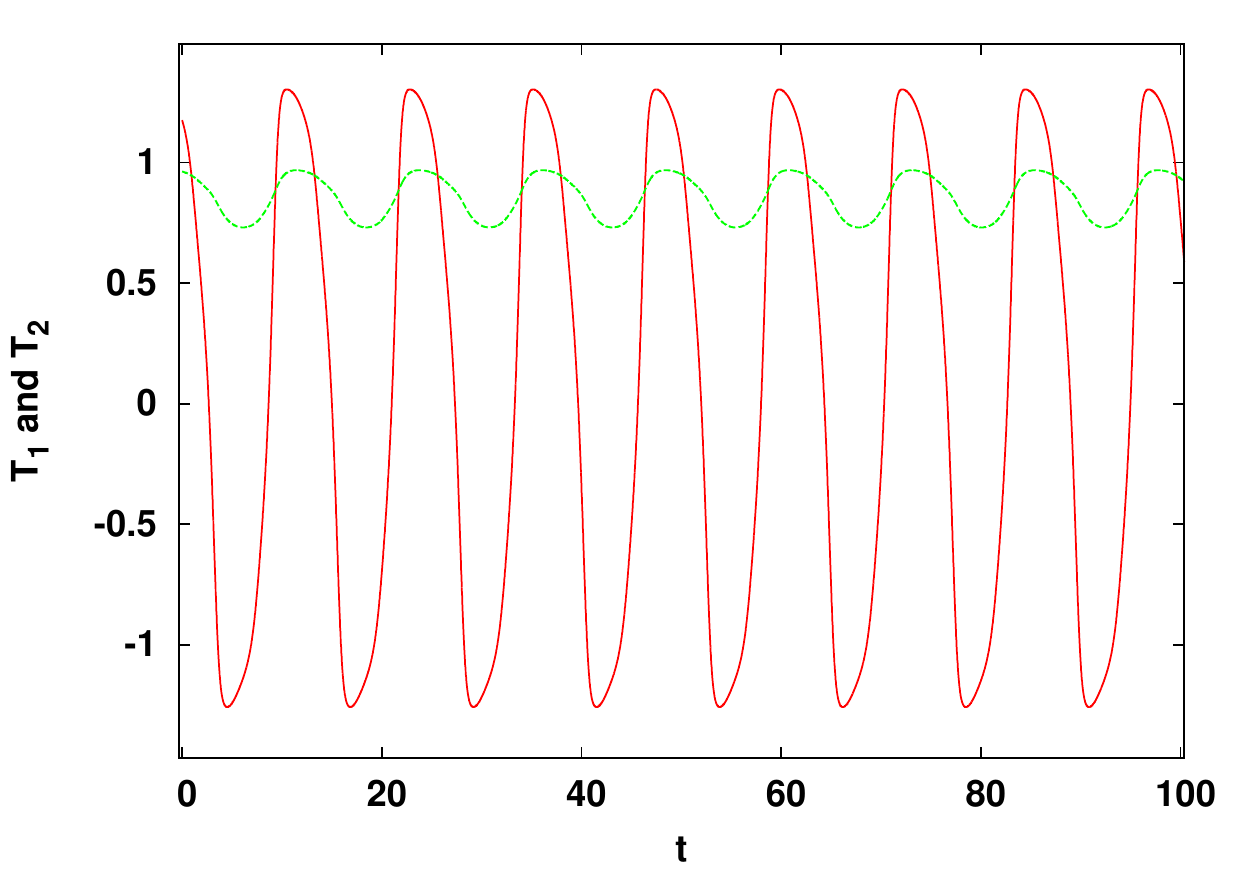}
			\includegraphics[scale=\SCALE]{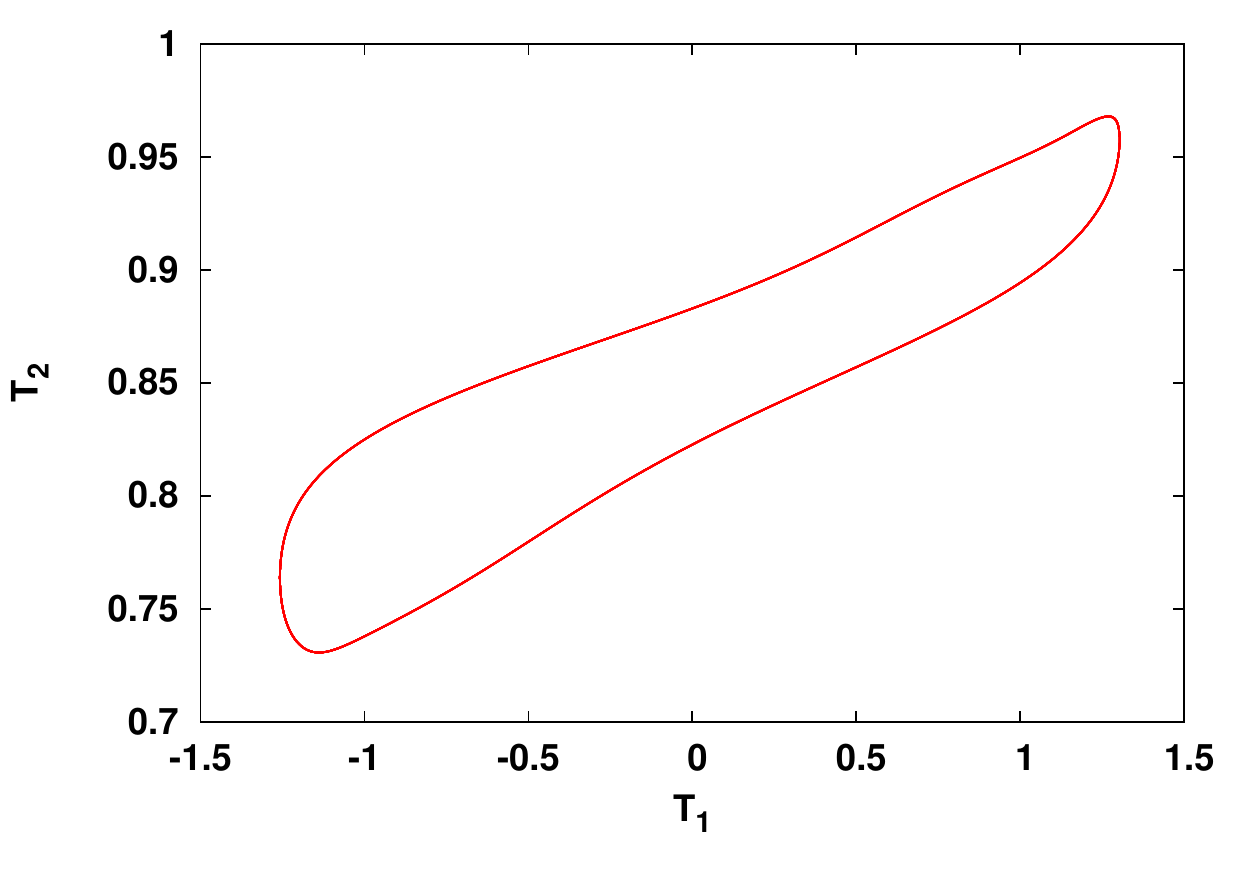}
			
			(a)
			\vspace{1cm}
			
		   \includegraphics[scale=\SCALE]{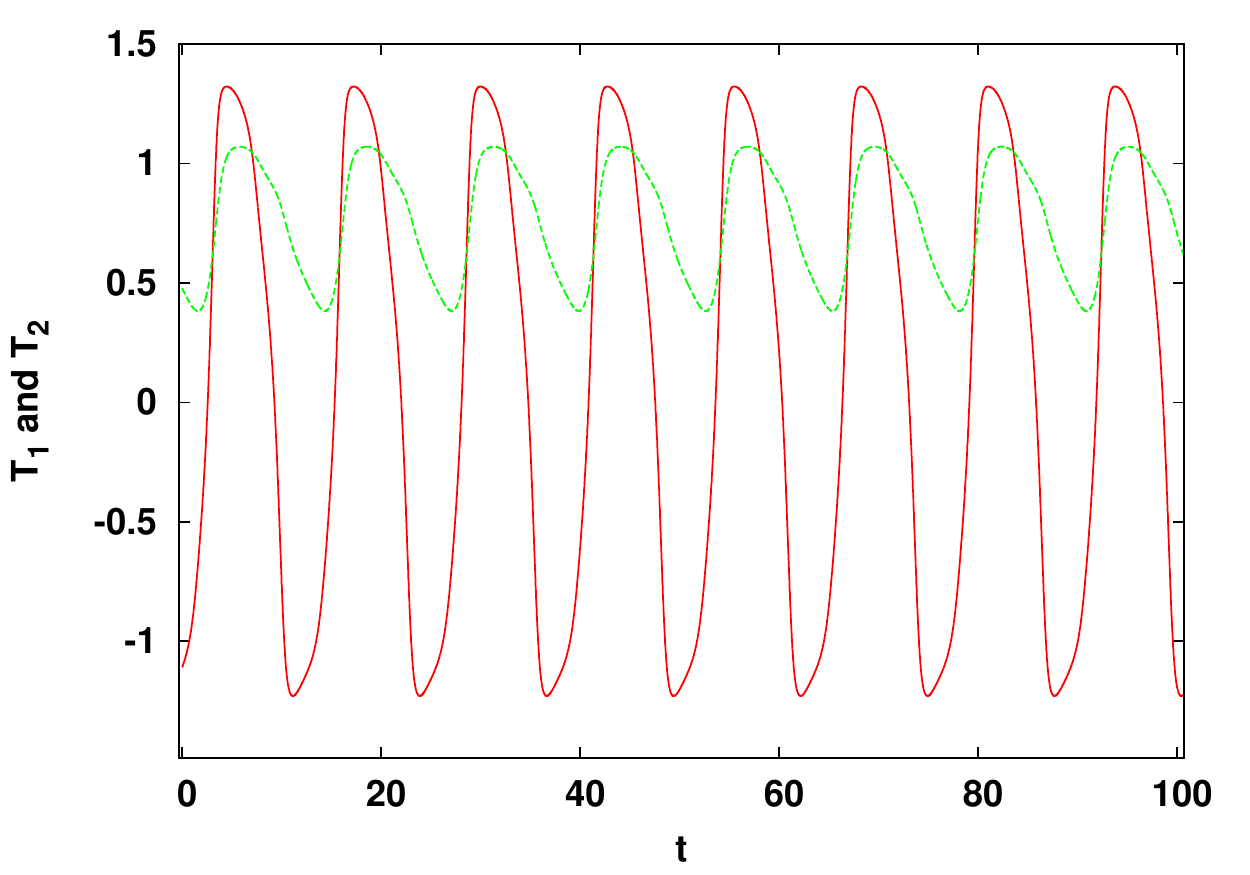} 
			\includegraphics[scale=\SCALE]{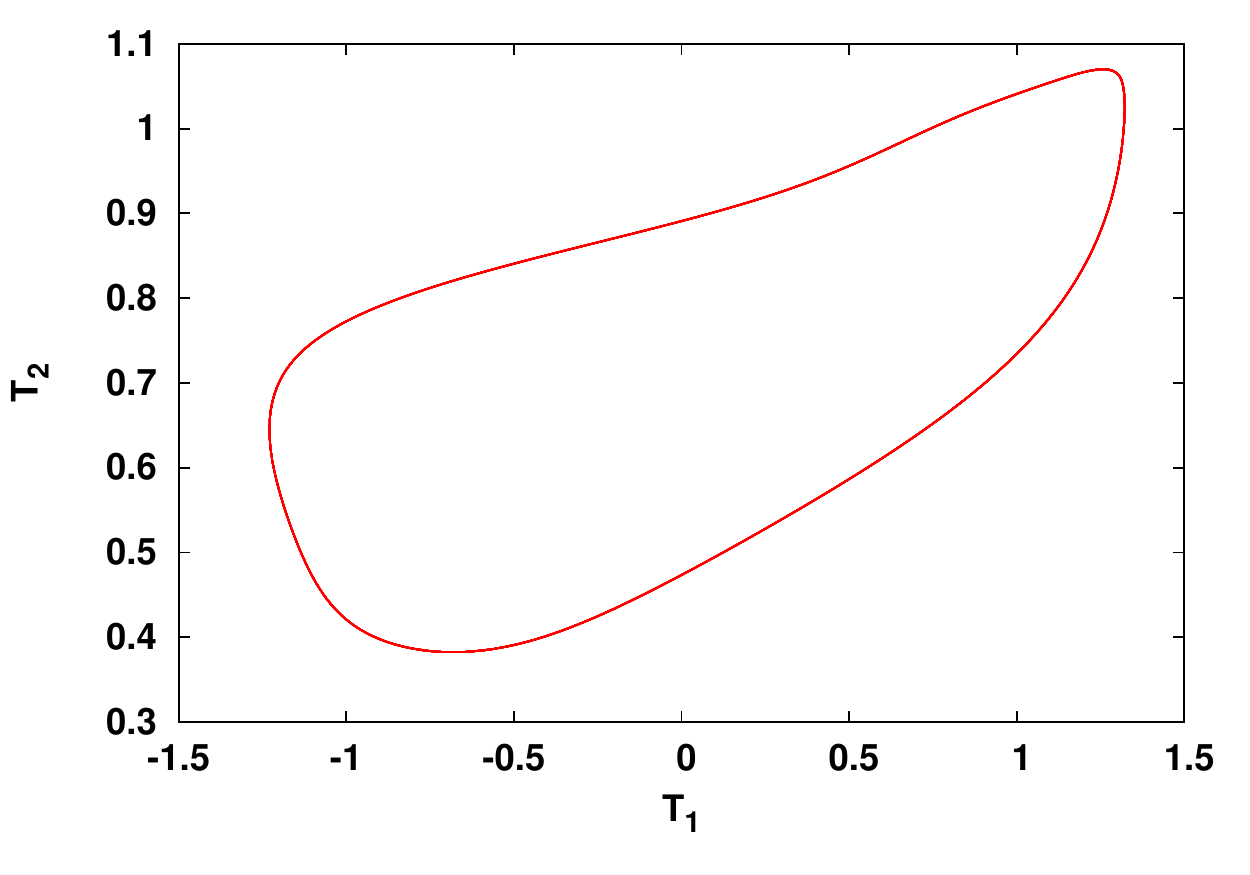}
			 
			(b)
			
			\caption{Temporal evolution of the temperature anomalies of the two sub-regions $T_1$ (in red) and $T_2$ (in green), in the left panels, and the corresponding phase portraits in the $T_1-T_2$ plane in the right panels, for a system with $\alpha_{1} = 0.75, \alpha_{2} = 0.25$, delay $\delta =4$ and inter-region coupling strength $\gamma$ equal to (a) $0.1$ and (b) $0.2$, in Eqn.\ref{main} \cite{csf_ms}.}\label{delta_alpha_2}		
		\end{figure}

\begin{figure}[H]
			\centering 
				\includegraphics[scale=\SCALE]{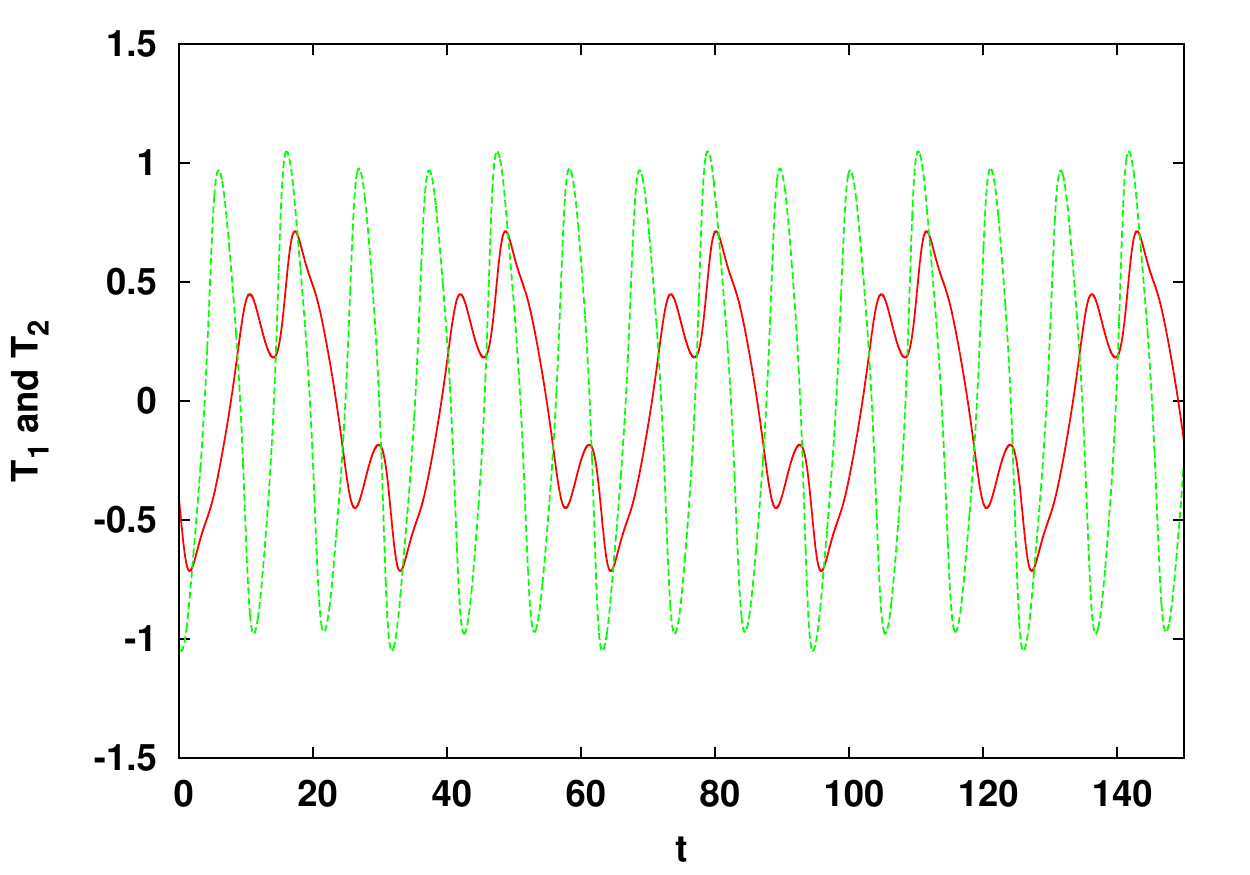}
				\includegraphics[scale=\SCALE]{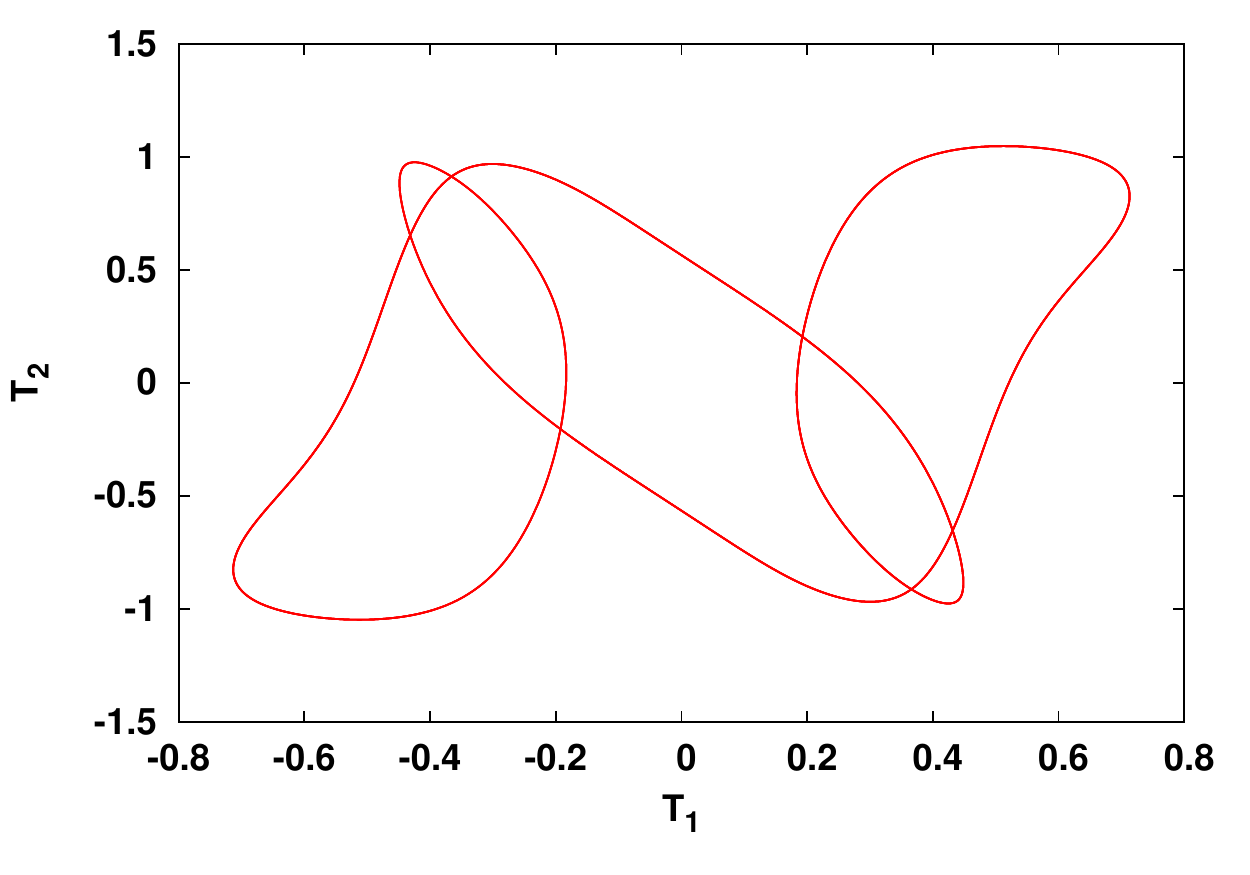} 
				
				(a)
				\vspace{1cm}

				\includegraphics[scale=\SCALE]{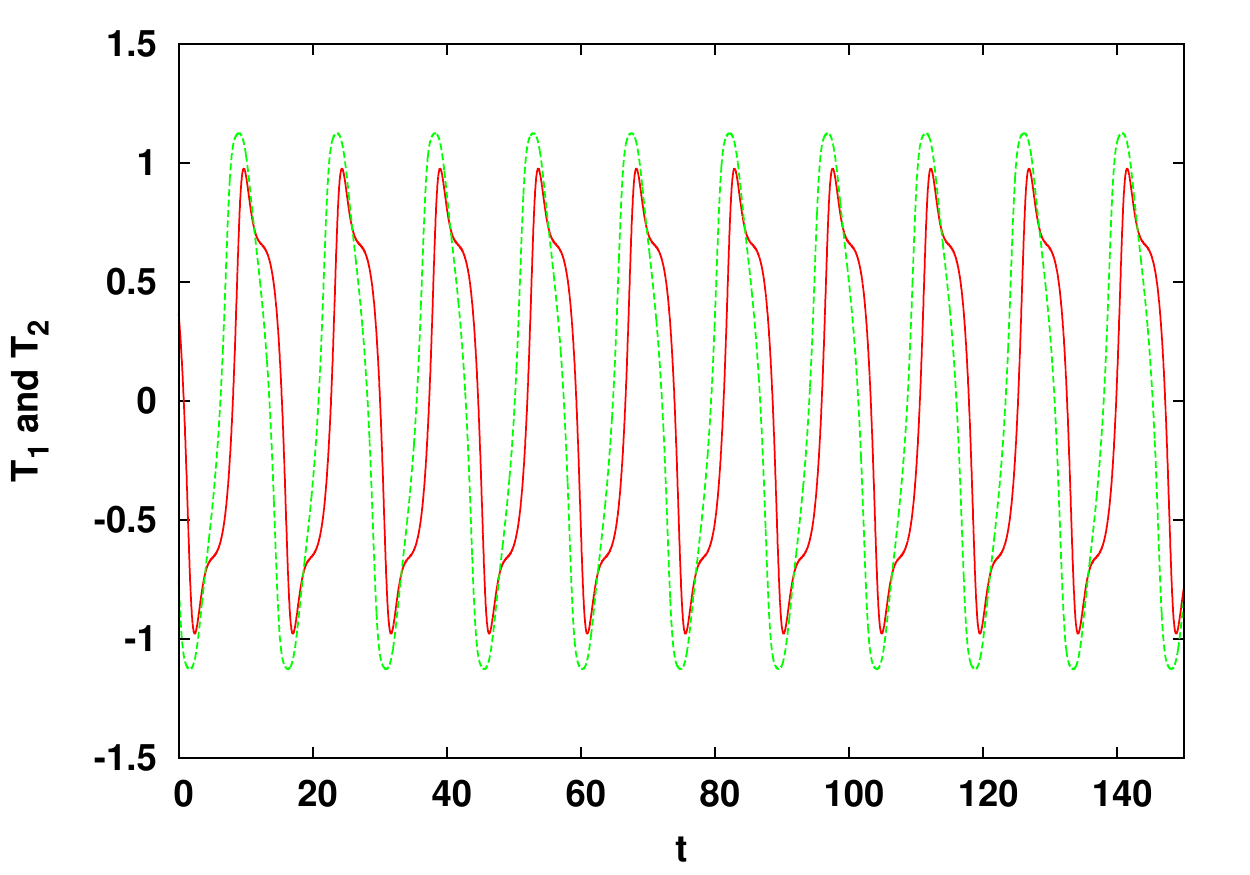}
				\includegraphics[scale=\SCALE]{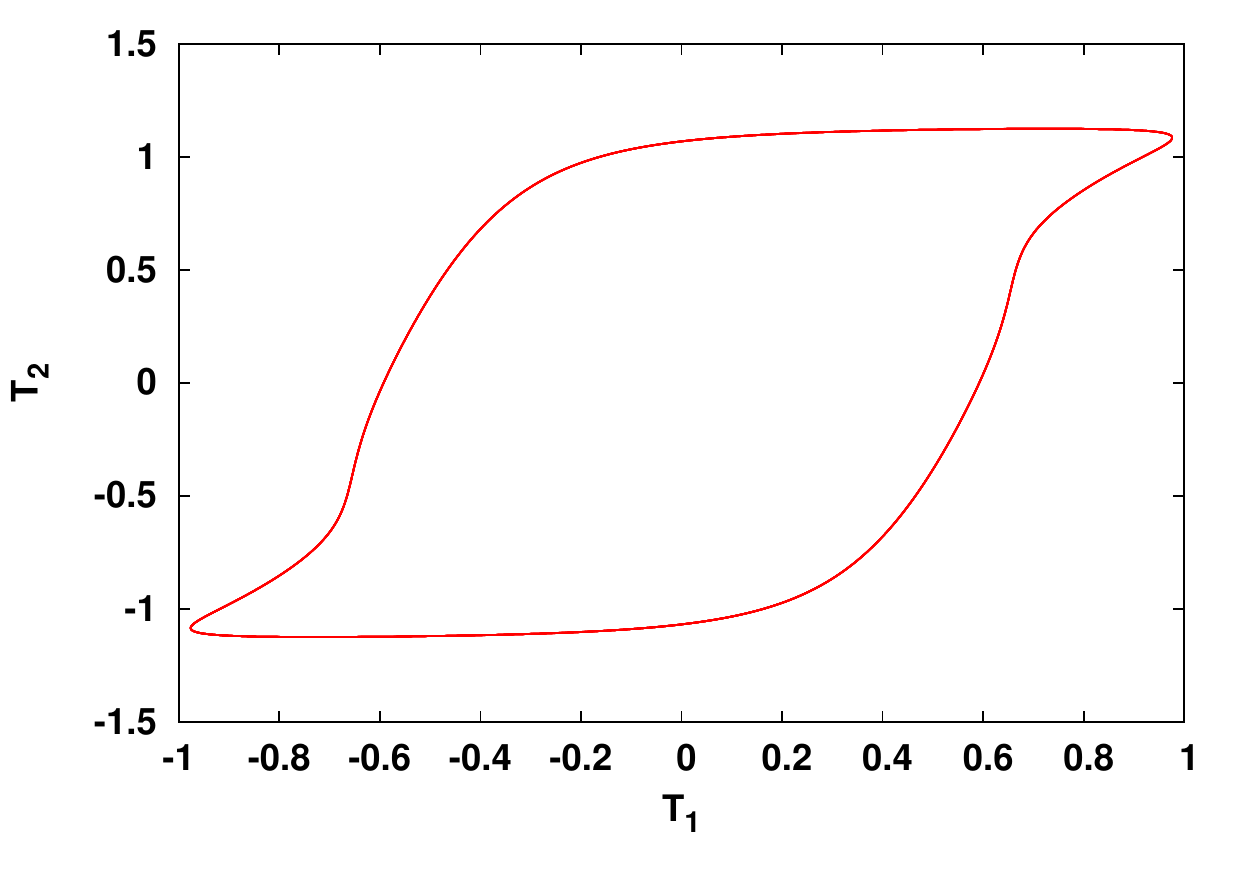}
				
				 (b)
				 \vspace{1cm}

				\includegraphics[scale=\SCALE]{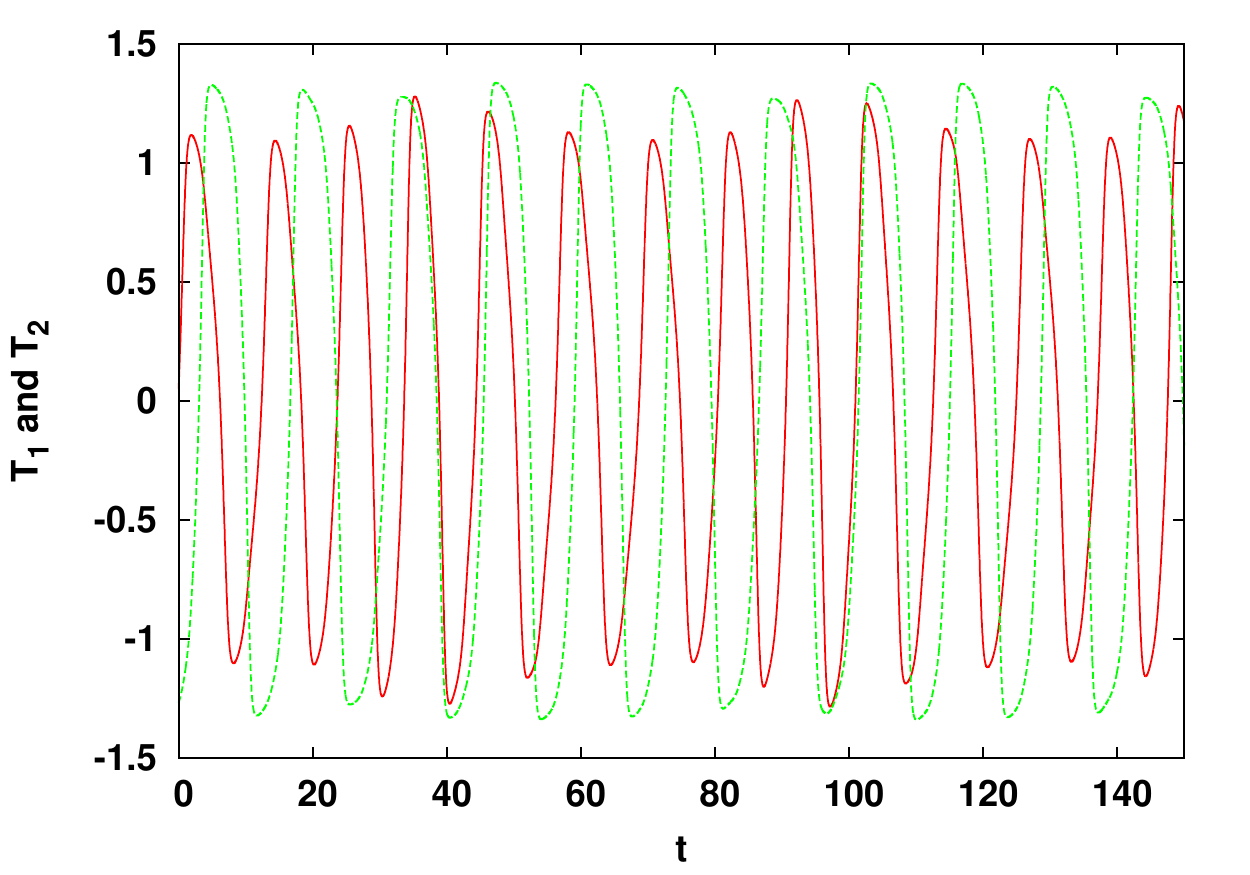}
				\includegraphics[scale=\SCALE]{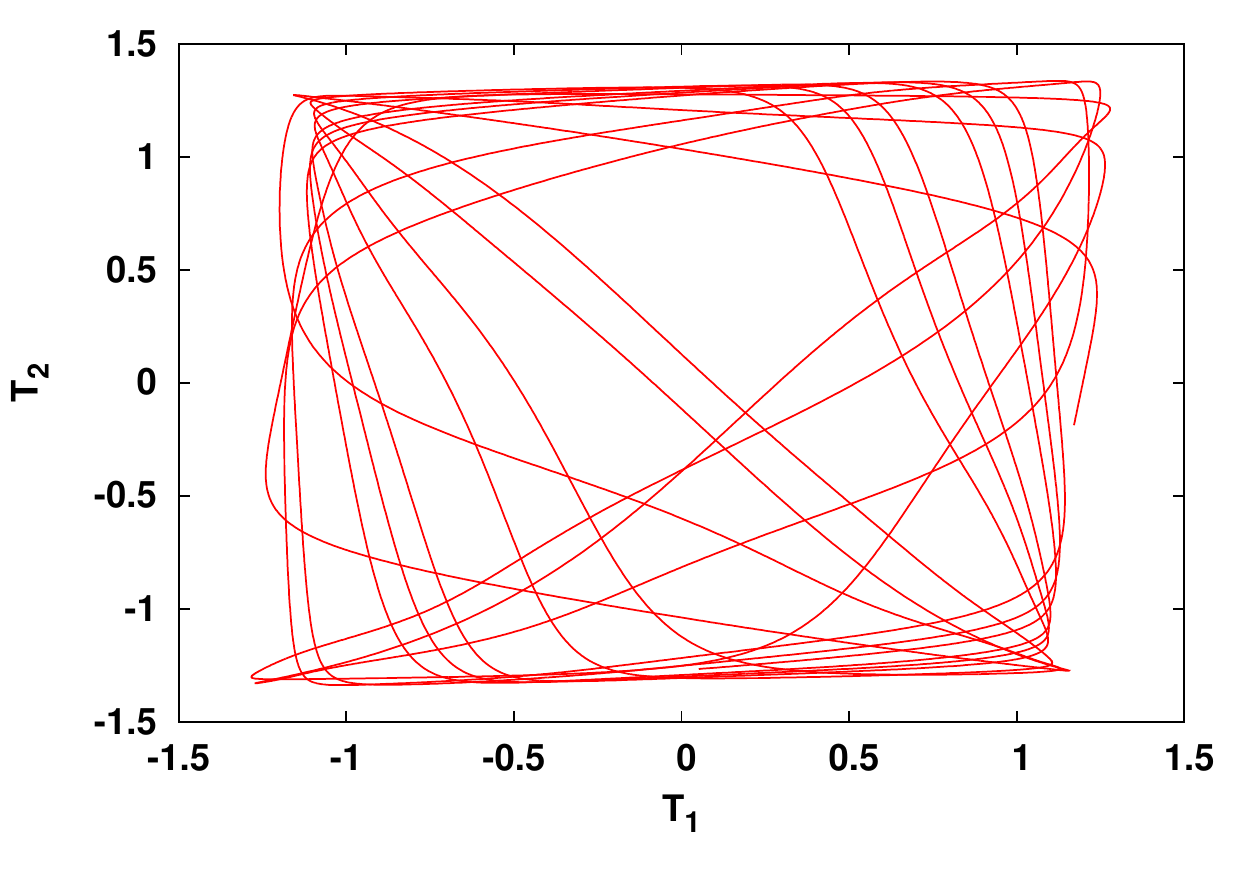} 
				
				(c)
			\caption{Time evolution of the temperature anomalies of the two sub-regions, with $\alpha_{1}=\alpha_{2}=0.75$ (a) $\delta_{1} = 1$, $\delta_{2} = 2$ and $\gamma=0.1$; (b) $\delta_{1} = 3$, $\delta_{2} = 5$ and $\gamma=0.1$; (c)  $\delta_{1} = 1$, $\delta_{2} = 3$ and $\gamma=0.3$;. The temperature anomaly of region 1, $T_{1}$, is shown in red and for region 2, $T_{2}$ is shown in green. The corresponding phase portrait is displayed on the right panel \cite{csf_ms}.}
			\label{patterns}
		\end{figure}

			When the difference in the strengths of the self-delay coupling  is small ($\Delta \alpha < \alpha_{1,2}$), we observe that both sub-regions display similar behaviour for strong inter-region coupling (cf. Fig. \ref{delta_alpha_1}b ). However for weaker inter-region coupling, different dynamical behaviour emerges in the two sub-regions. Typically, the region with stronger self-delay coupling shows regular behaviour, while the region with weaker self-delay coupling shows complex behaviour. This type of complex oscillation is qualitatively very similar to ENSO observational data \cite{Soon2,basicmodel}.\\\\
			When the difference in $\alpha$ is large ($\Delta \alpha > \alpha_{1,2}$), then the nature of oscillations in the two sub-regions can be very different. For instance in Fig. \ref{delta_alpha_2} one observes that one sub-region displays large amplitude oscillations in the temperature anomaly, while the other sub-region displays very small amplitude oscillations. So we see that non-uniformity in the self-coupling strengths in the systems can significantly affect the temperature anomaly of mean sea surface temperature in neighbouring sub-regions.  Further, it was clearly seen that the {\em parameter region supporting oscillations is larger for weaker inter-region coupling strengths and small difference in self-delay coupling strengths of the two sub-regions}.\\\\
			When the self-delays are different, with $\delta_{1} \ne \delta_{2}$, complex oscillatory patterns arise. These complex patterns are also qualitatively similar to the actual observations of the ENSO phenomena. Representative examples of these are shown in Fig. \ref{patterns}.\\\\
			We also estimated the basin of attraction for the fixed point state, by finding the fraction of initial conditions that evolve to fixed points. If this fraction is one, the fixed point state is the global attractor of the dynamics. When this fraction is zero, none of the sampled initial conditions evolve to fixed points, and the system goes to an oscillatory state instead. When the fraction is larger than zero and less than one, we have co-existence of attractors (namely certain initial conditions evolve to fixed points, while others yield oscillations).\\\\
			The estimated basin of attraction clearly showed that the region of co-existence of fixed points and oscillations is narrower for lower inter-region coupling, and wider for higher inter-region coupling strengths. Thus it is a evident that strong inter-region coupling $\gamma$ favours larger parameter regions of oscillation suppression, and also yields a larger parameter range where fixed points states co-exist with oscillatory states.

\begin{figure}[H]
				\centering 
				\includegraphics[scale=\SCALE]{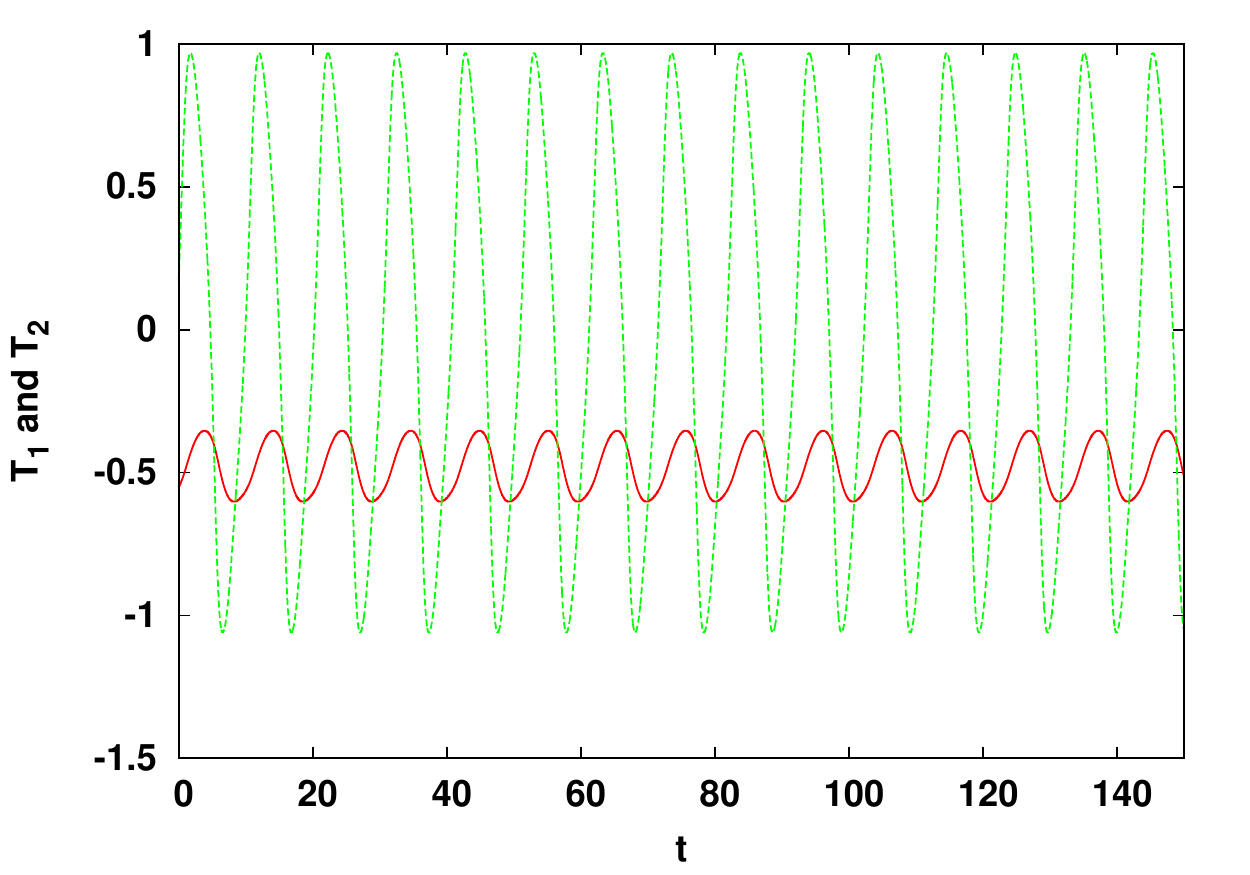}
				\caption{Time evolution of the temperature anomalies of the two sub-regions, with self-delay $\delta_{1} = 0$ in region 1 and $\delta_{2} = 2$ in region 2. The inter-region coupling strength is $\gamma = 0.1$ and self-delay coupling strength is $\alpha=0.75$. The temperature anomaly of region 1, $T_{1}$, is shown in red and for region 2, $T_{2}$ is shown in green  \cite{csf_ms}.}\label{fp1}
		\end{figure}

		Interestingly, if we take the self-delay coupling strengths of the two sub-regions to be such that the temperature of one region goes to a fixed point regime when uncoupled, while the other system is in the oscillatory regime, then on coupling both systems show oscillations (see Fig. \ref{fp1}). This implies that oscillations may arise in certain sub-regions through coupling to neighbouring regions. Namely, a sub-region with very low delay ($\delta<2$), which would naturally go to a steady state when uncoupled, yields oscillations when coupled to  another sub-region with high enough delay ($\delta>2$).

	\section{Basins of attraction of the different emergent dynamical states}

\begin{figure}[H]	
				\centering 
				\includegraphics[scale=\SCALE]{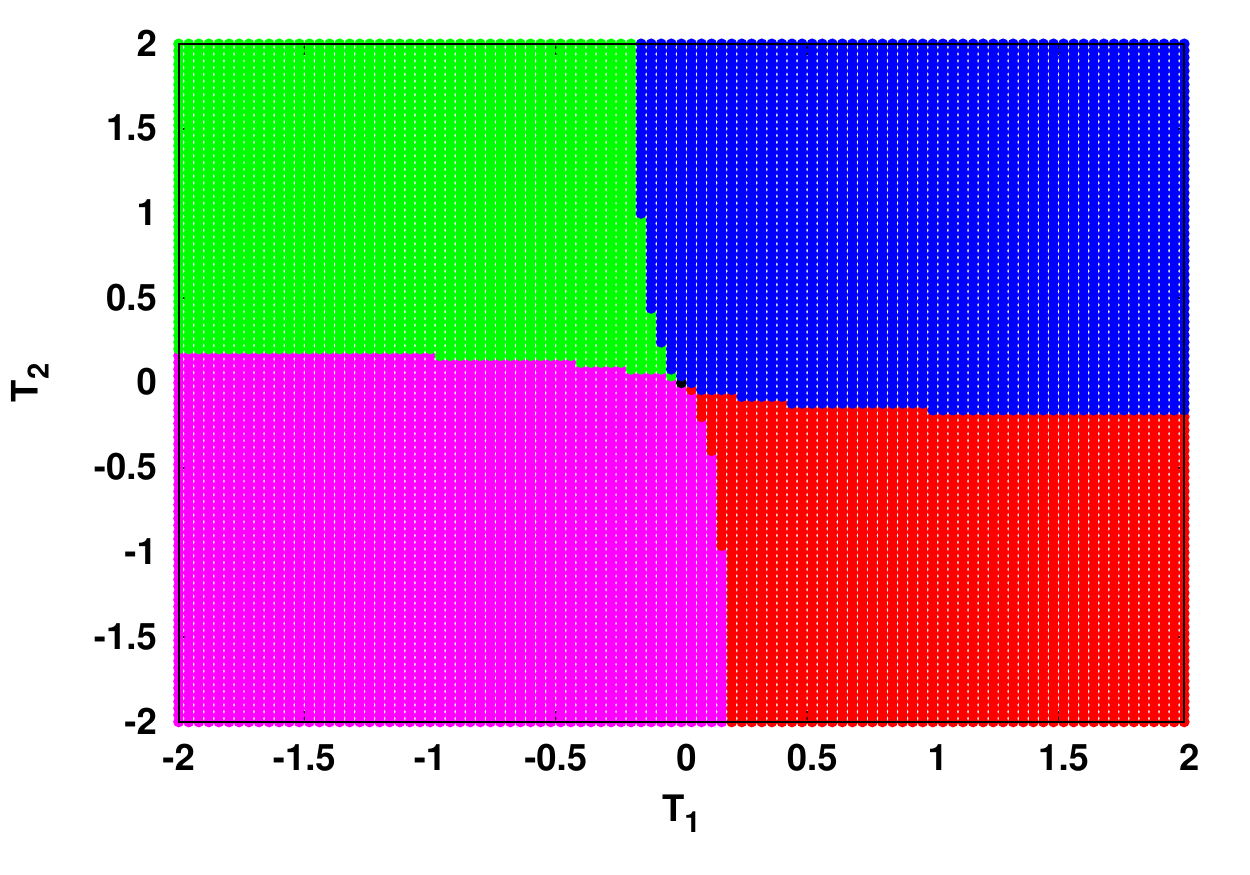}
				\includegraphics[scale=\SCALE]{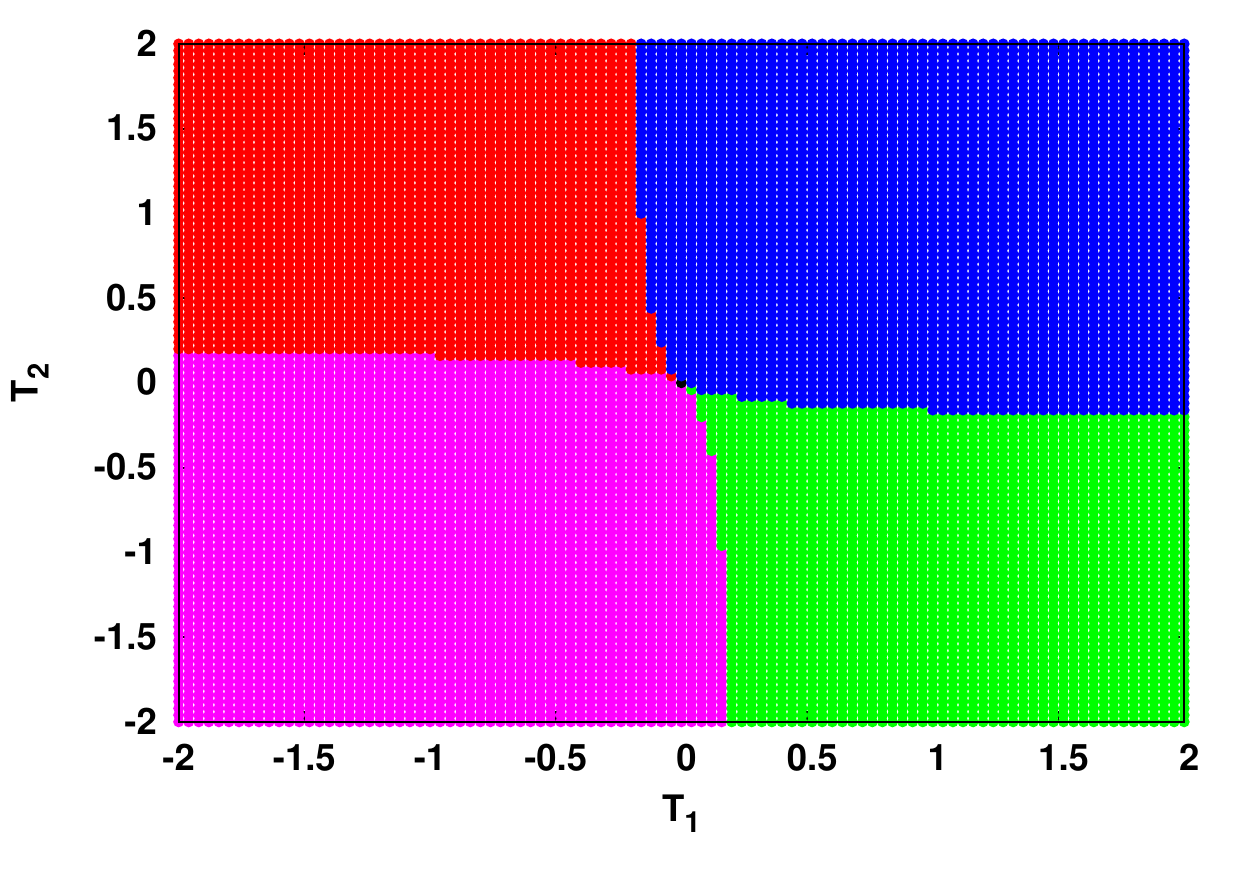}
				
				(a)
				\vspace{1cm}
				
				\includegraphics[width=\twoFigureSize]{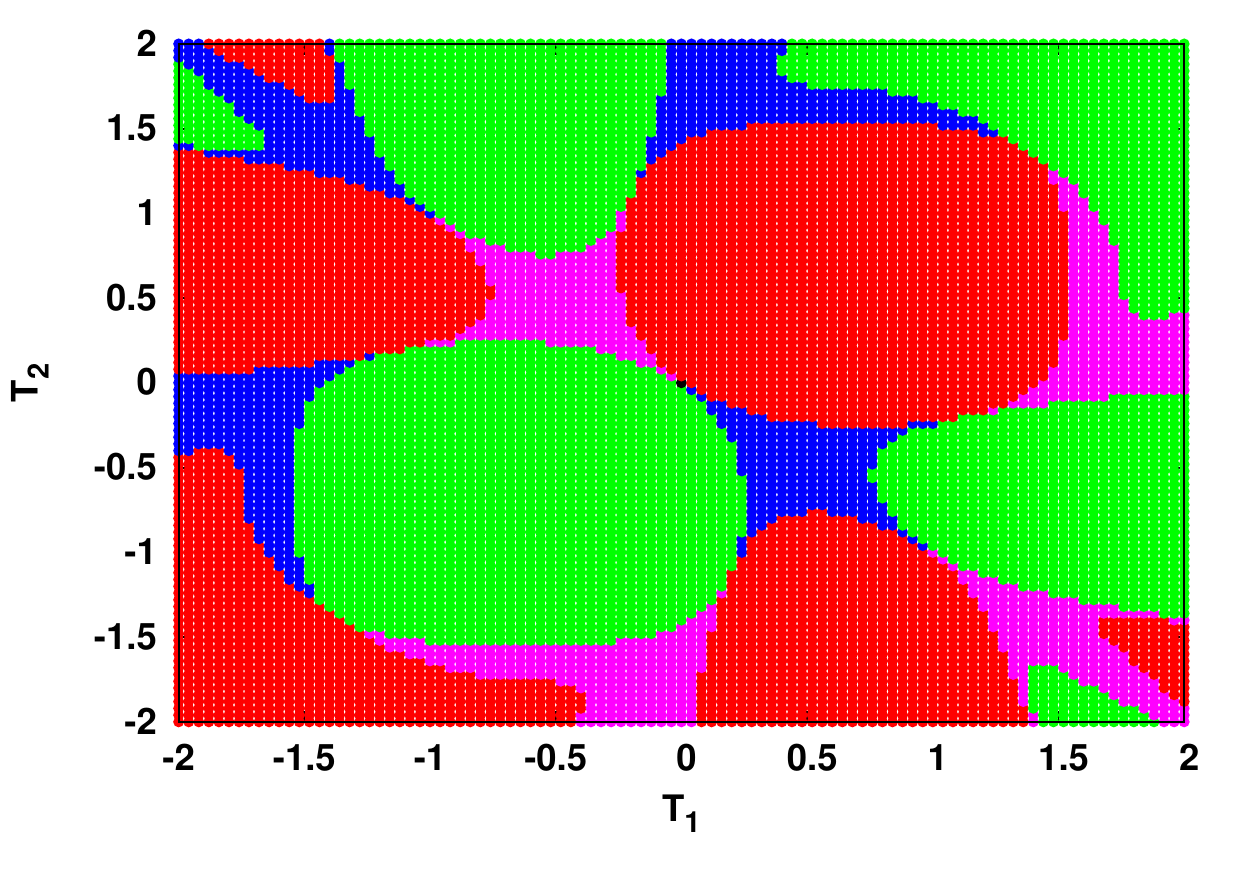}
				\includegraphics[width=\twoFigureSize]{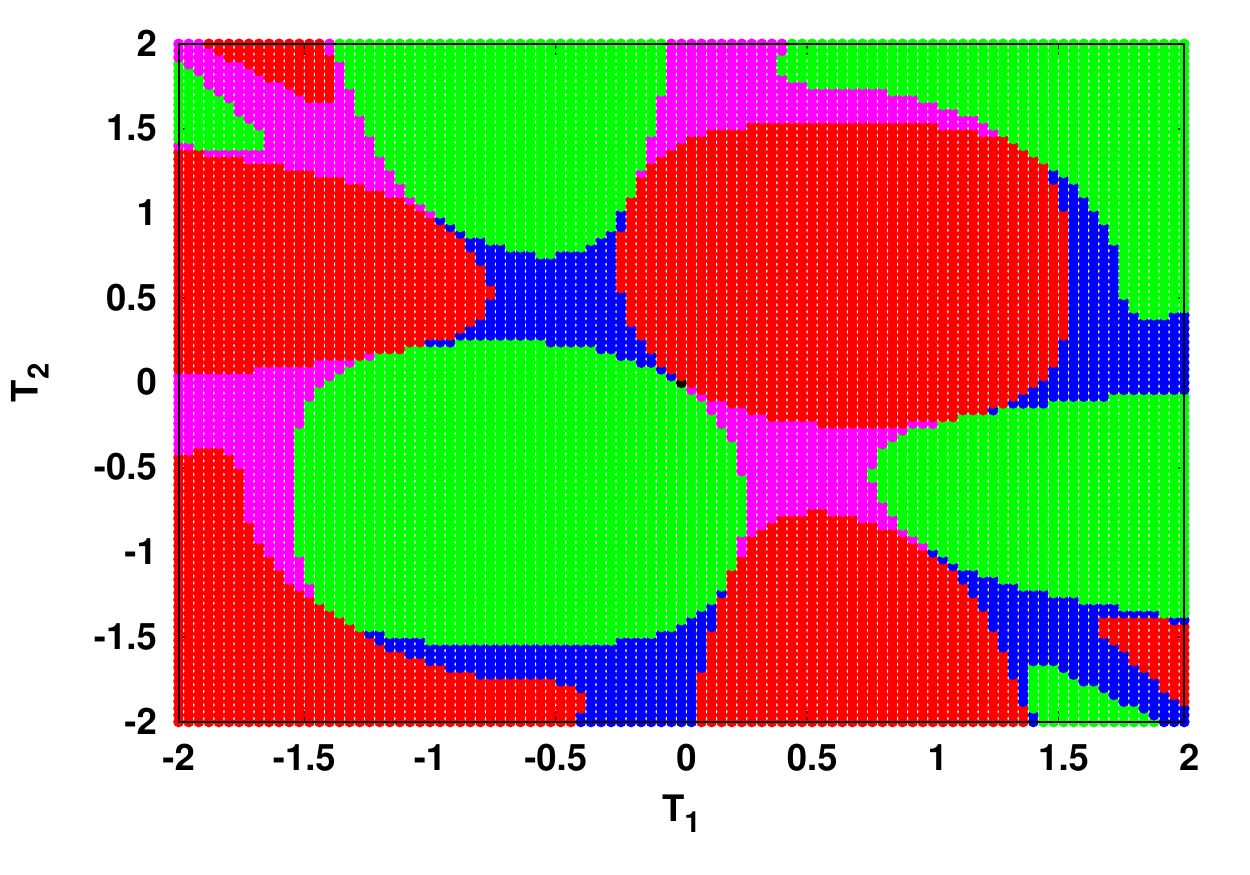}
				
			    (b)
			    
			    \caption{Basins of attraction of the different dynamical attractors, in the space of scaled temperature anomalies $T_1$ and $T_2$. Here the green, red, magenta and blue colors represent the basins of attraction of fixed point attractors. The system parameters are $\delta_1=\delta_2=\delta=1, \gamma=0.1$ and (a) $\alpha_{1}=\alpha_{2}=0.5$ and (b) $\alpha_{1}=\alpha_{2}=\alpha=0.75$. The left panel is for sub-region 1 and the right panel shows sub-region 2.}\label{a_0p75_g=0P1}
		\end{figure}

\begin{figure}[H]
				\centering 
				\includegraphics[scale=\SCALE]{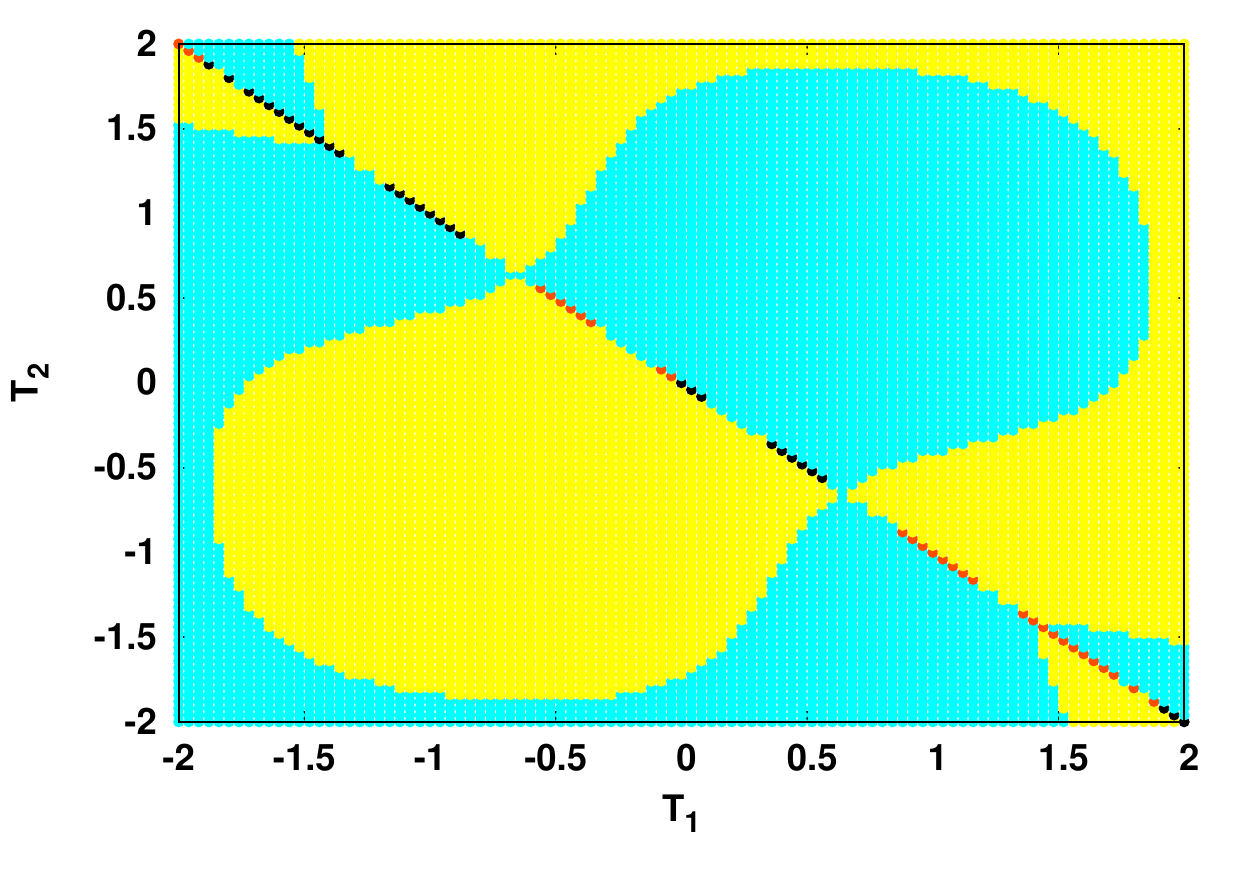}
				\includegraphics[scale=\SCALE]{bs_T1_d=1_g=0p2_a=0p75}
				
				(a)
			
				\includegraphics[scale=\SCALE]{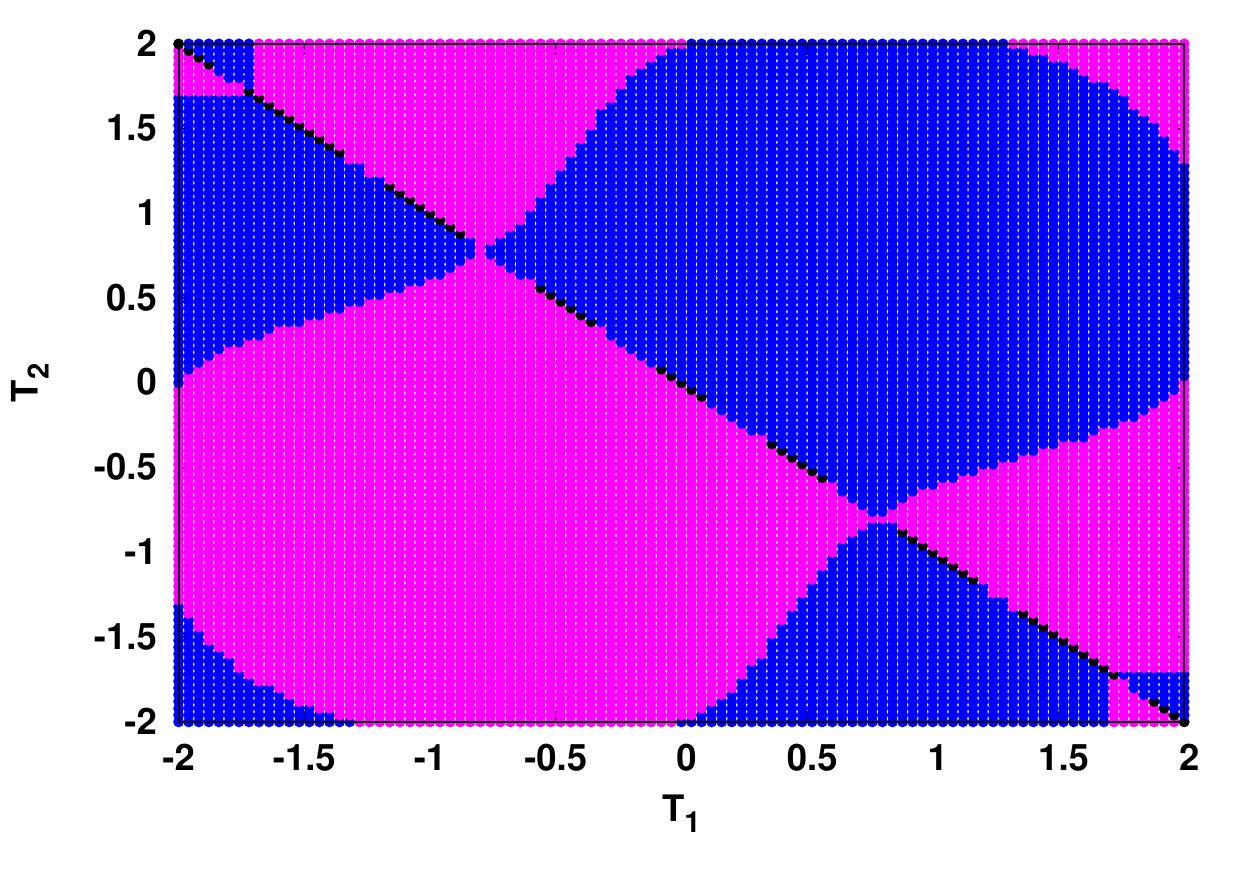}
				\includegraphics[scale=\SCALE]{bs_T1_d=1_g=0p3_a=0p75}

				(b)
			    
				\includegraphics[scale=\SCALE]{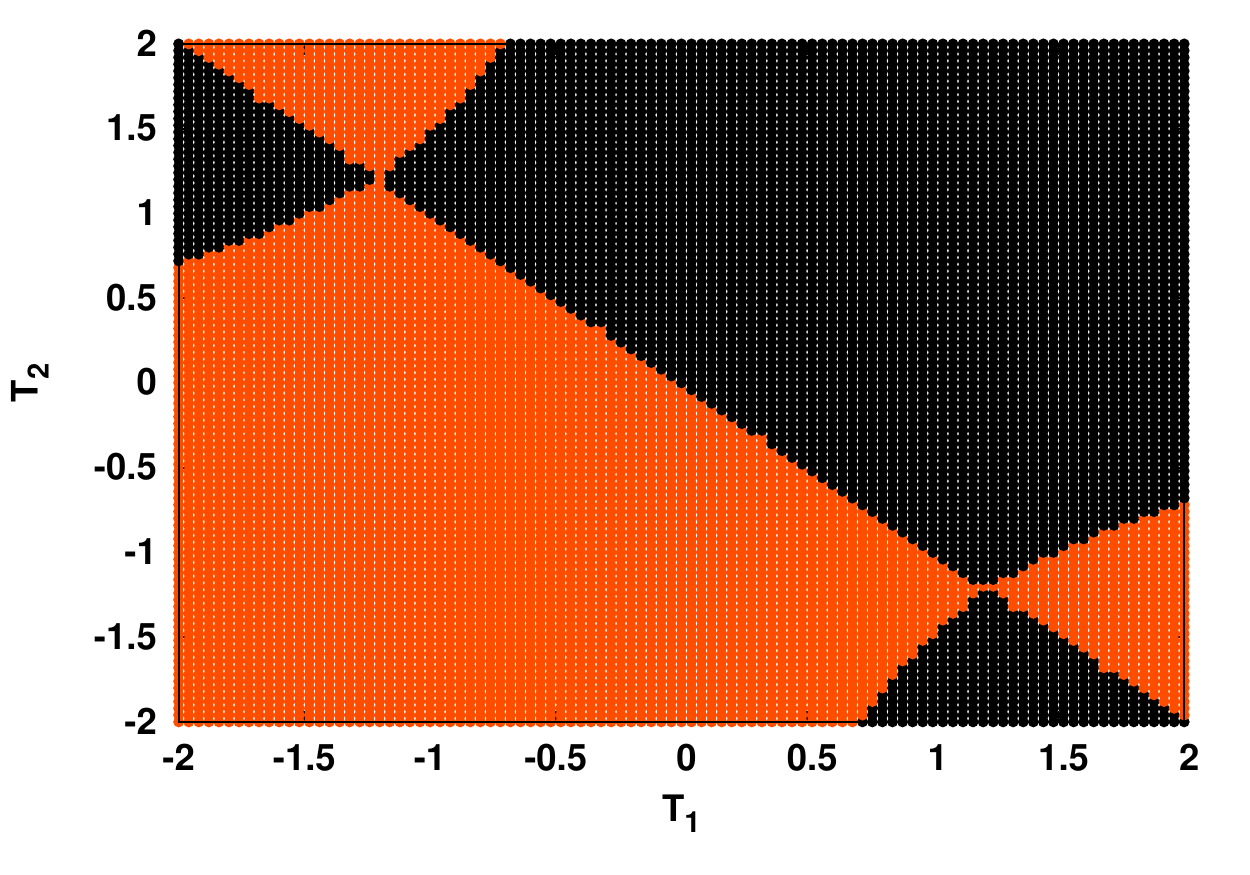}
				\includegraphics[scale=\SCALE]{bs_T1_d=1_g=0p7_a=0p75}

				(c)
				
				\caption{Basins of attraction of the different dynamical attractors, in the space of scaled temperature anomalies $T_1$ and $T_2$. Here the yellow, light blue, magenta, blue and black and orange colors represent the basins of attraction of a fixed point attractor. The system parameters are $\alpha_{1}=\alpha_{2}=\alpha=0.75, \delta_1=\delta_2 = \delta=1$, and (a) $\gamma=0.2$, (b) $\gamma=0.3$ and (c) $\gamma=0.7$. The left panel is for sub-region 1 and the right panel shows sub-region 2.}
\label{a_0p75_g=0P2}
						
		\end{figure}

\begin{figure}[H]
			\centering 
			\includegraphics[scale=\SCALE]{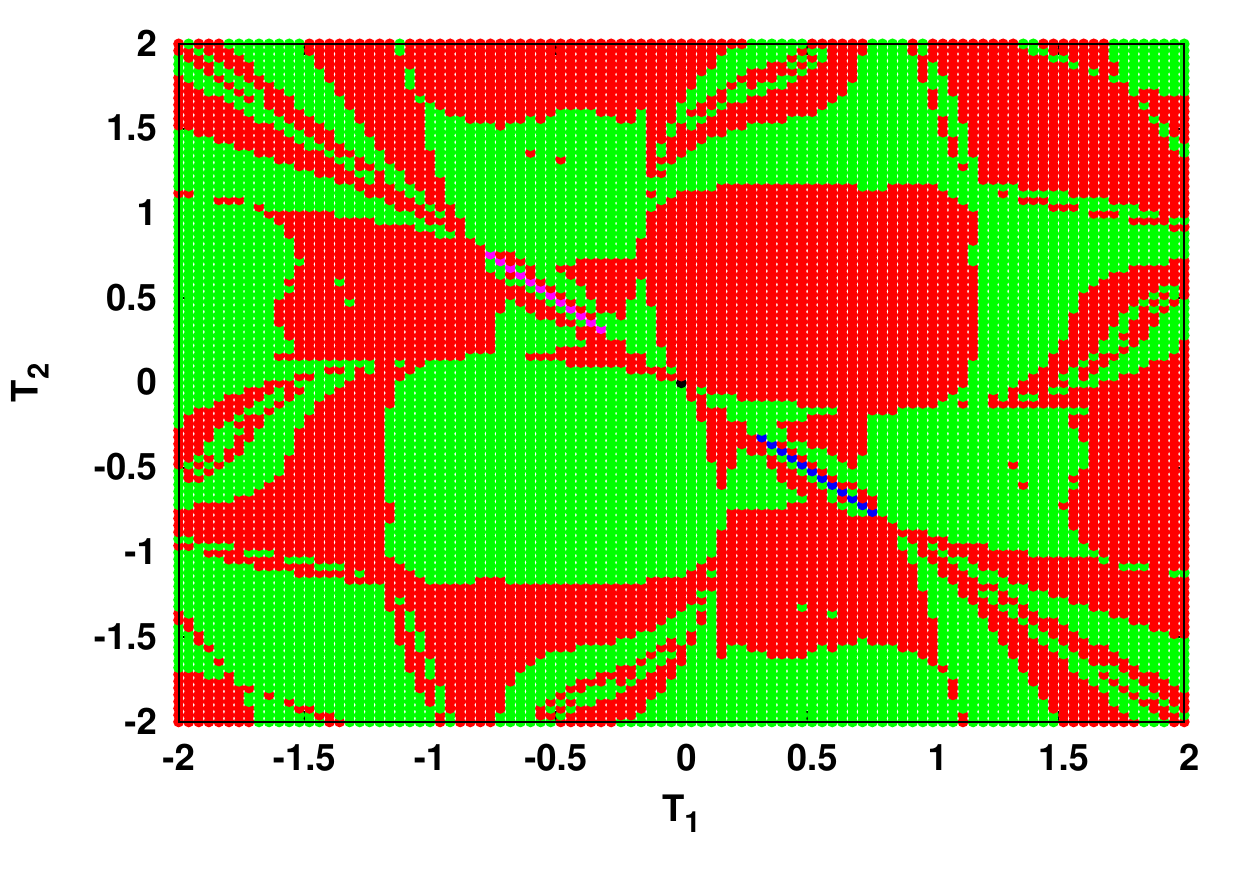}
			\includegraphics[scale=\SCALE]{bs_T1_d=4_g=0p1_a1=0p5_a2=0p5}
			\caption{Basins of attraction of the different dynamical attractors, in the space of scaled temperature anomalies $T_1$ and $T_2$. Here the green and red colors represent the basins of attraction of fixed point attractors. The system parameters are:  $\alpha_{1}= \alpha_{2}=\alpha=0.5, \delta_1=\delta_2=\delta=4, \gamma=0.1$. The left panel is for sub-region 1 and the right panel shows sub-region 2.}
			\label{a1_p5_g_p1_d=4}
			
			\vspace{2cm}
			
			\includegraphics[scale=\SCALE]{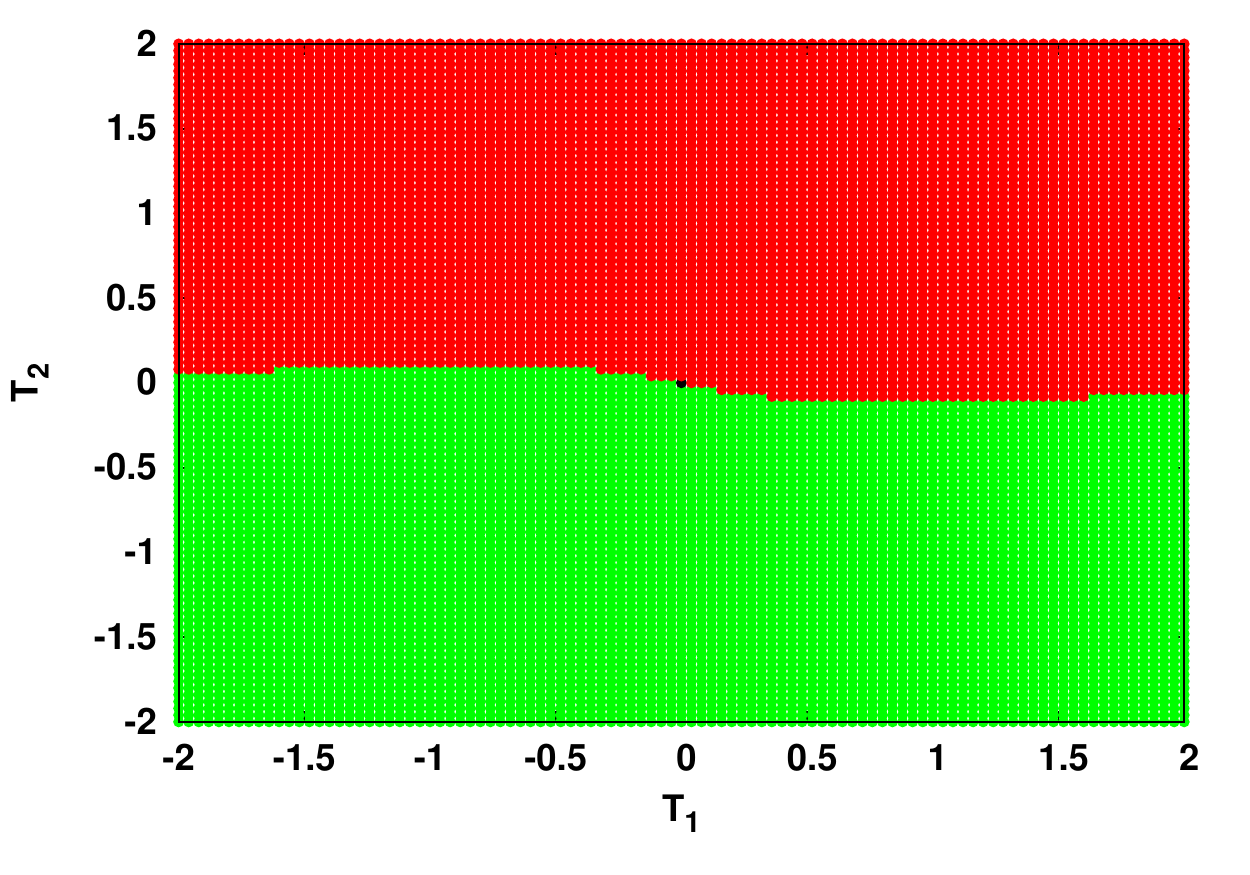}
			\includegraphics[scale=\SCALE]{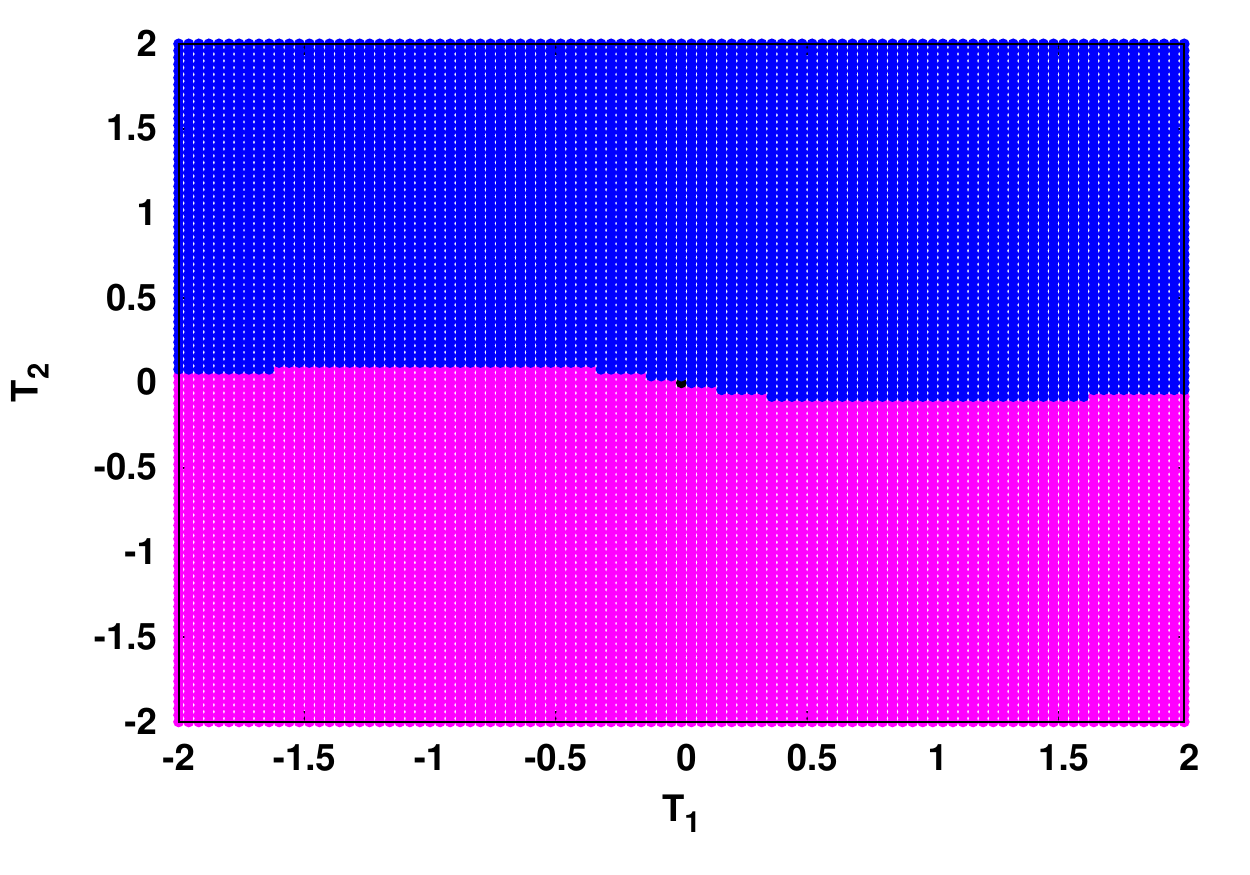}
			\caption{Basins of attraction of the different dynamical attractors, in the space of scaled temperature anomalies $T_1$ and $T_2$. Here the green, red, magenta and blue colors represent the basins of attraction of fixed point attractors.  The system parameters are: $\alpha_{1}=0.75 , \alpha_{2}=0.5, \delta_1=\delta_2=\delta=1, \gamma=0.1$. The left panel is for sub-region 1 and the right panel shows sub-region 2.}\label{a1_p75_g_p1_d_1}
			\end{figure}

	The basin of attraction of a dynamical state is the set of points in the space of the system variables, such that if the initial conditions are chosen in this set, the system will evolve to that particular state. In our model we have observed many different dynamical attractors, ranging from fixed points to low-amplitude and high amplitude oscillations. The number of co-existing attractors and their basins of attraction depend crucially upon the self delay, delay and inter-region coupling strengths. So the estimation of the basins of attraction of the different states is important here, as it indicates the prevalence of the state in general and the probability of observing the state given a window of initial conditions. We present below representative cases of the different emergent states and their basins of attraction, for the case of coupled identical sub-regions, as well as coupled non-identical sub-regions.\\
	We first examine the case of coupled identical sub-regions. Specifically we present the case of delays $\delta_1 = \delta_2 =1$, for different inter-region coupling strengths $\gamma$. When the inter-region coupling is weak, for instance the case of $\gamma=0.1$ displayed in Fig.~\ref{a_0p75_g=0P1}, both sub-regions have four distinct steady states. For the case of $\alpha_{1}=\alpha_{2}=0.5$ (Fig.~\ref{a_0p75_g=0P1}a), when the initial values of the sub-regions are both positive or both negative, then both sub-regions approach the same steady state. However, when the initial states are different, namely one region is positive and the other negative, then they approach different steady states, i.e. one positive and one negative steady state. So two dynamical attractors have the same basins of attraction in the sub-regions, while the other two attractors have different basins of attraction, with the basins of the two states being switched. Similarly for the case of $\alpha_{1}=\alpha_{2}=0.75$ (Fig.~\ref{a_0p75_g=0P1}b), we find that two of the fixed-point attractors have same basin of attraction in $T_1-T_2$ space in the two sub-systems, while the basins of attraction of the other two fixed-point attractors is switched in the two sub-systems. Interestingly, now the fixed-point attractors which have same basin of attraction in both sub-regions, have larger basin volume as compared to the two attractors that have different basins of attraction in the sub-systems. \\
When $\gamma \ge 0.2$  (cf. Fig.~\ref{a_0p75_g=0P2}) we obtain two steady states, one of which is a positive fixed point and the other a negative fixed point. The positive fixed point state is bounded entirely in a window of positive values (as represented by the light blue, blue and black colors) and negative fixed point state is bounded entirely in a window of negative values (as represented by the yellow, magenta and orange colors). The specific values of the fixed points depend on the values of the inter-region coupling strength $\gamma$. For a particular value of $\gamma$, the basin of attraction for each attractor is same in both sub-systems and we observe an inversion symmetry of the attractors along the diagonal.\\ Further, for large delays, for instance the case of $\delta_1=\delta_2=4$ displayed in Fig.~\ref{a1_p5_g_p1_d=4}, it is clear that the sub-regions yield two attractors of the same type, with the same basins of attraction in the sub-regions. However, {\em as the delay increases, the basin boundaries become very complex.}

			\begin{figure}[H]
			\centering 
			\includegraphics[scale=\SCALE]{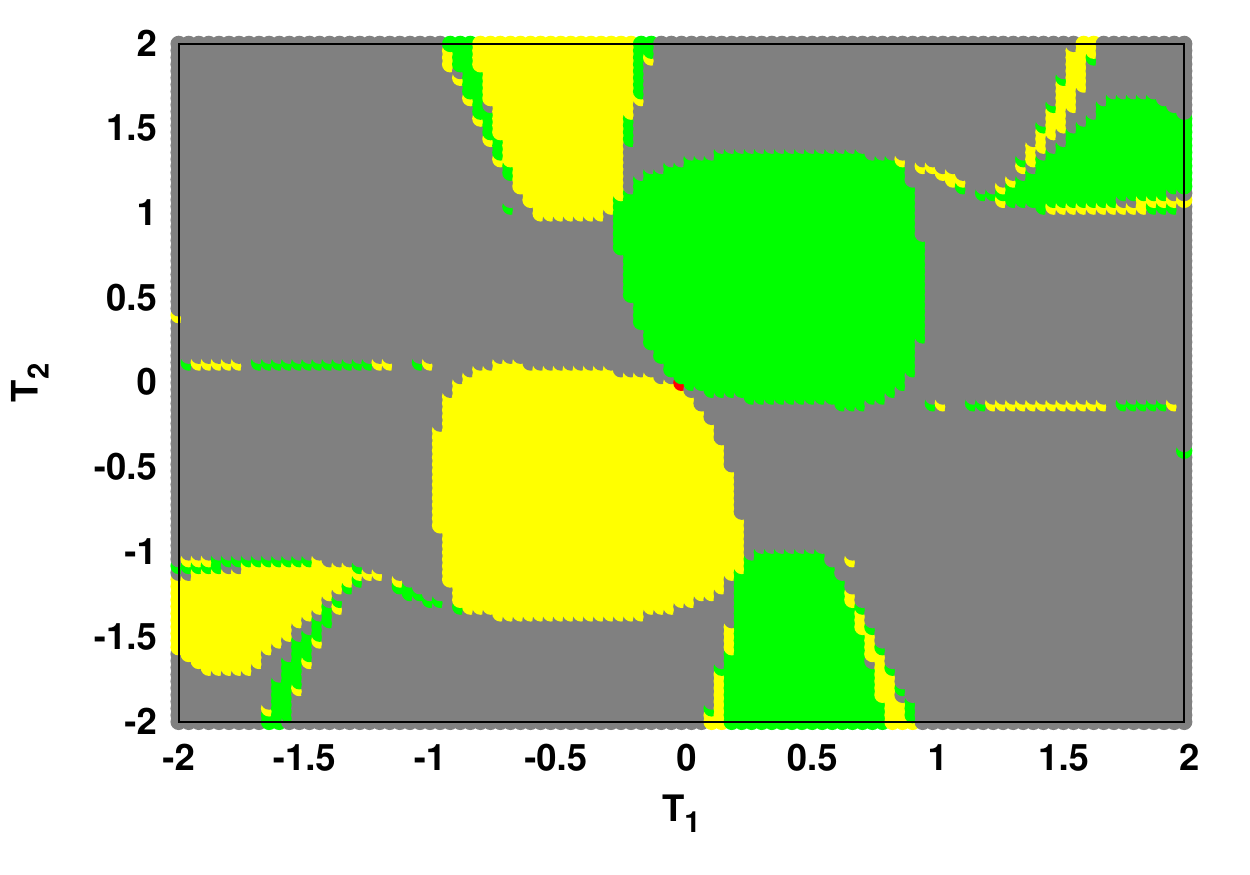}
			\includegraphics[scale=\SCALE]{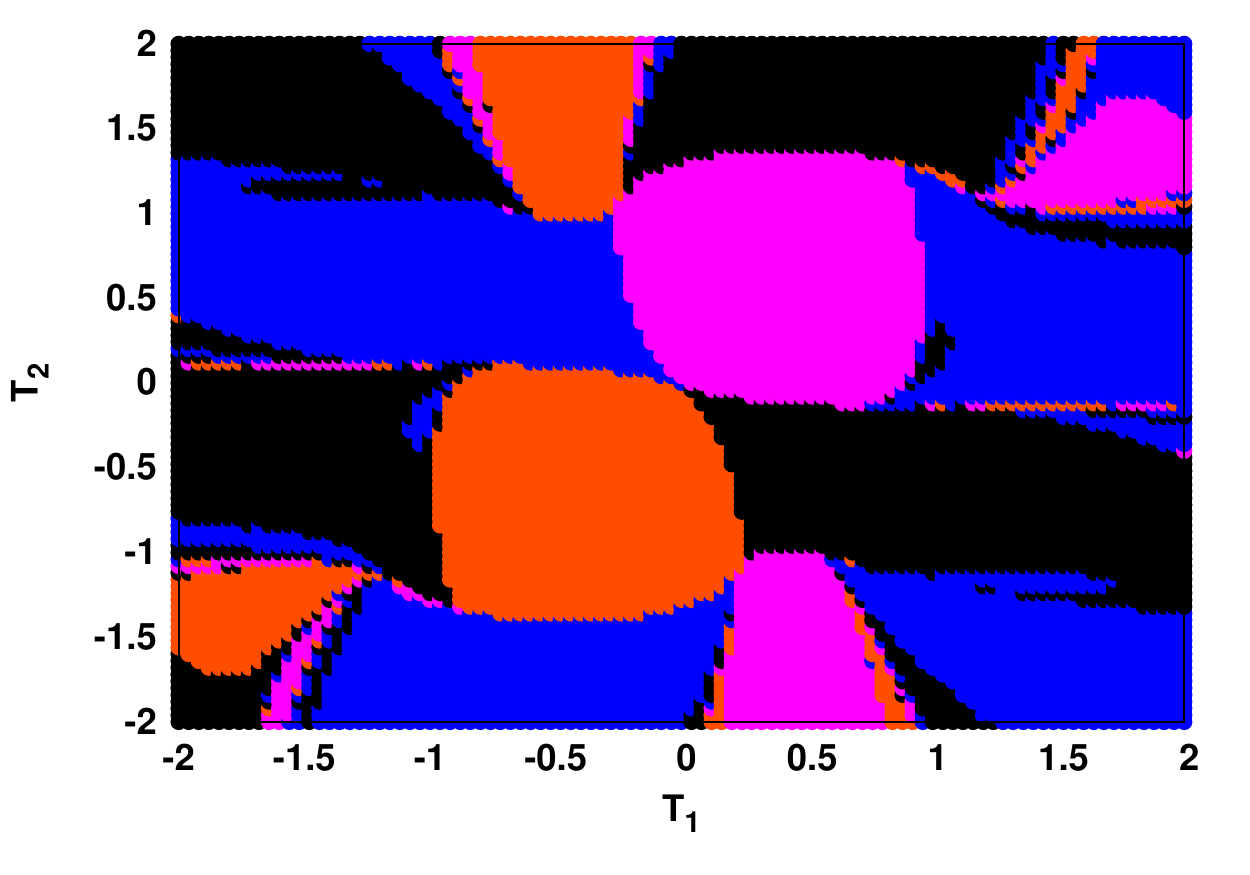}

(a)

		\includegraphics[scale=\SCALE]{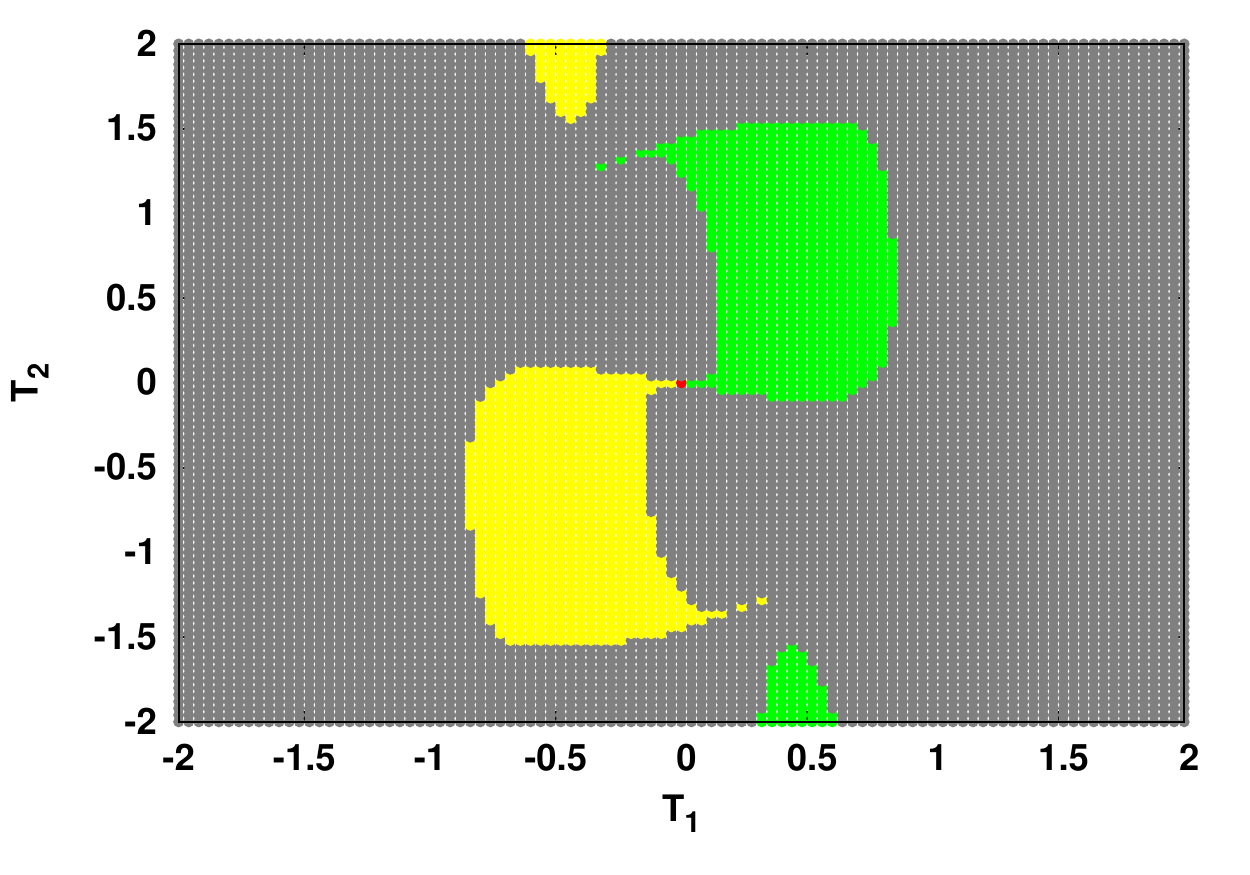}
		\includegraphics[scale=\SCALE]{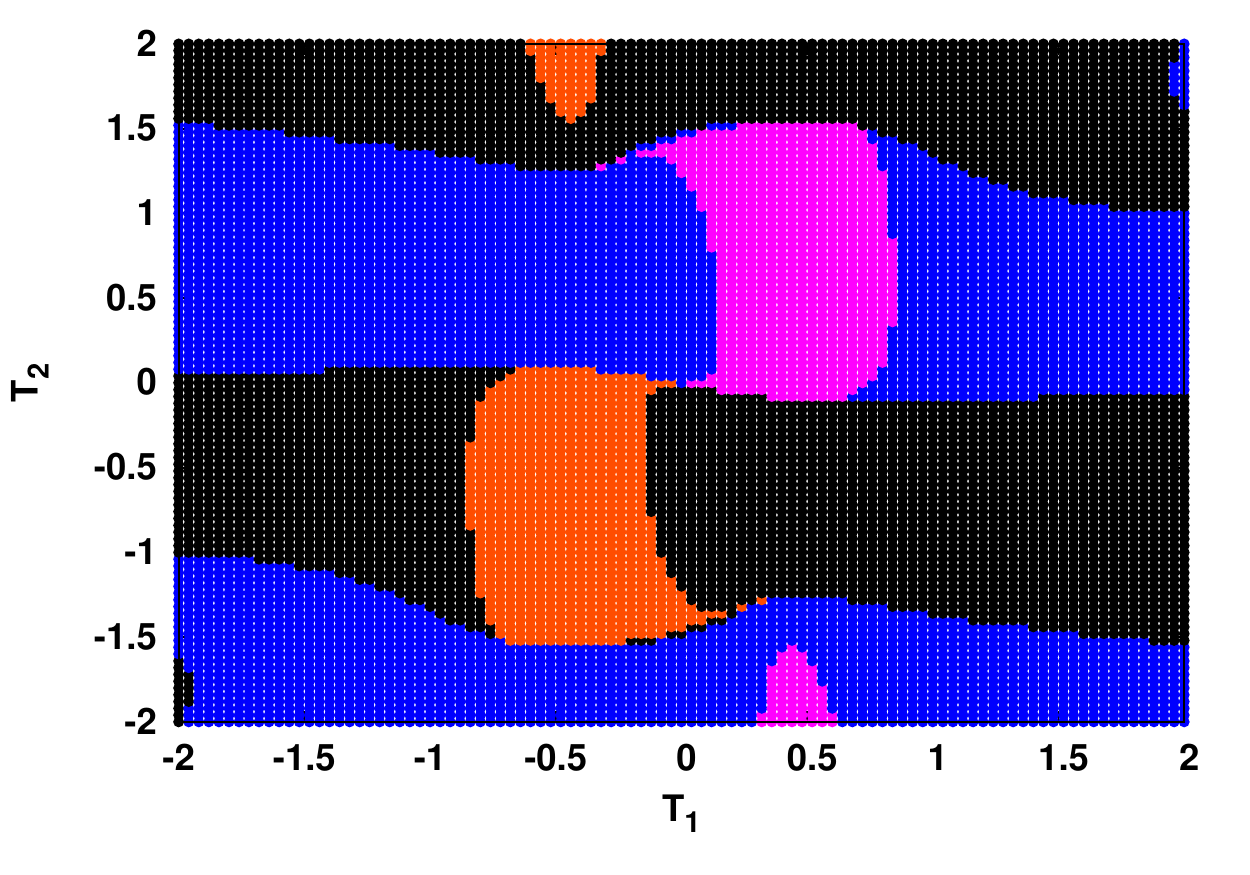}

(b)

			\caption{Basins of attraction of the different dynamical attractors, in the space of scaled temperature anomalies $T_1$ and $T_2$. Here the green, yellow, magenta and orange colors represent the basins of attraction of fixed point attractors. The black and blue colors represent the basins of attraction of low amplitude oscillations. For the blue color, the minimum and maximum values of the oscillations are positive and for the black color the minimum and maximum values of the oscillations are negative. The gray color represents large amplitude oscillations, with positive value of maxima and negative value of minima. The system parameters are: (a) $\alpha_{1}=0.65, \alpha_{2}=0.5, \delta_1=\delta_2=\delta=2.5, \gamma=0.1$; (b) $\alpha_{1}=0.75, \alpha_{2}=0.5, \delta_1=\delta_2=\delta=2, \gamma=0.1$. The left panel is for sub-region 1 and the right panel shows sub-region 2.}\label{fp_os_a1_p65}
		\end{figure}

		   Next we consider the case of coupled non-identical systems. For instance, we display an illustrative example of a system with parameters $\alpha_{1}=0.75$, $\alpha_{2}=0.5$, $\gamma=0.1$ and $\delta=1$ in Fig. \ref{a1_p75_g_p1_d_1}. Here we observe two types of attractors, and these attractors are different in the two sub-regions due to the difference in the value of the self-delay coupling strengths. Specifically, for small delays the dynamical attractors are fixed points, and the values of these fixed points are dependent on the values of the parameters. As the delay increases we observe coexistence of fixed points and oscillatory attractors (cf. Fig. \ref{fp_os_a1_p65}a-b). 
				   
	   From the case of $\alpha_{1}=0.75, \alpha_{2}=0.5, \delta=2, \gamma=0.1$, displayed in Fig. \ref{fp_os_a1_p65}b, we observe that there are three types of attractors for the system when the strength of self delay is high. Two of these attractors are fixed points (represented by green and yellow colors) and one is an oscillatory attractor with high amplitude (represented by the gray color). Systems with weak self-delay strength have four types of attractors, two of which are fixed points (represented by magenta and orange color) and two are low-amplitude oscillatory attractors, that are entirely positive-valued or negative-valued. We also observe that for large delay as strength of self-delay increases, the volume of the basin of attraction of the fixed point attractors decreases.
	   
\section{Robustness of the dynamical attractors under noise} 

If the system is attracted to a dynamical state, even under the influence of noise, then the state can be considered robust under noise. In order to examine this, we examine the system described by Eqn.~\ref{main}, under Gaussian noise:
                \begin{eqnarray} 
                           \label{main_noise}
                                        \frac{dT_{1}}{dt}= T_{1}-T_{1}^3-\alpha_{1} T_{1}(t-\delta_{1})+\gamma T_{2} + D \eta(t)\\ 
\nonumber
                                        \frac{dT_{2}}{dt}= T_{2}-T_{2}^3-\alpha_{2} T_{2}(t-\delta_{2})+\gamma T_{1} + D \eta(t)
                \end{eqnarray} 
where $\eta$ is a delta-correlated Gaussian noise and $D$ is the strength of the noise. Here both sub-systems experience the same noise. 

First, consider a system with identical strengths of self-delay $\alpha_{1}=\alpha_{2}=0.75$, delay $\delta_1=\delta_2=\delta=1$ and inter-region coupling $\gamma=0.1$, where there are four fixed point solutions for the noise-free system: $0.591608$, $-0.591608$, $0.38729$ and $-0.38729$. Now to check the robustness of the different fixed points, we add noise to the system, and follow the evolution of the noisy system from different initial values of $T_{1}$ and $T_2$. Specifically, in Fig.~\ref{a_0p75_ic=p_0p5}, the initial values of $T_{1}$ and $T_2$ is $0.5$. Without noise both sub-systems go to the fixed point at $0.591608$ (cf. Fig. {\ref{a_0p75_ic=p_0p5}a). Under weak perturbations this fixed point is still attractive, with the noisy system confined around the fixed point at $0.591608$ for low noise strengths (cf. Fig. {\ref{a_0p75_ic=p_0p5}b-c). However, interestingly, when the noise strength is high, the system {\em switches between the two states} around $0.591608$ and $-0.591608$ (cf. Fig.~\ref{a_0p75_ic=p_0p5}d). The system does not wander to the other two fixed points at $0.38729$ and $-0.38729$ at all, but only jumps randomly between two bands around $0.591608$ and $-0.591608$. 

Similarly, for the same system evolving from initial condition $T_1=0.5, T_2=-0.5$, it is evident from Fig.~\ref{a_0p75_ic=np_0p5} that each sub-system goes to different states $0.38729$ and $-0.38729$ when there is no noise (cf. Fig.\ref{a_0p75_ic=np_0p5}a). However, when noise strength is low (e.g. $D=0.01, 0.05$), the noisy system goes to either a state around $0.591608$ or around $-0.591608$, with both sub-systems now approaching the same state (cf. Fig.\ref{a_0p75_ic=np_0p5}b-c). So even under weak noise the system evolves away from the fixed points $0.38729$ and $-0.38729$, and is attracted to states around the fixed points at $0.591608$ and $-0.591608$. When noise strength is high, again there is {\em switching} between these states (cf. Fig.\ref{a_0p75_ic=np_0p5}d). Thus Figs.~\ref{a_0p75_ic=p_0p5}-\ref{a_0p75_ic=np_0p5} suggest that the fixed points $0.591608$ and $-0.591608$ are more robust to noise than the other two fixed points.

                \begin{figure}[H]
                	\centering 
                		\includegraphics[scale=\SCALE]{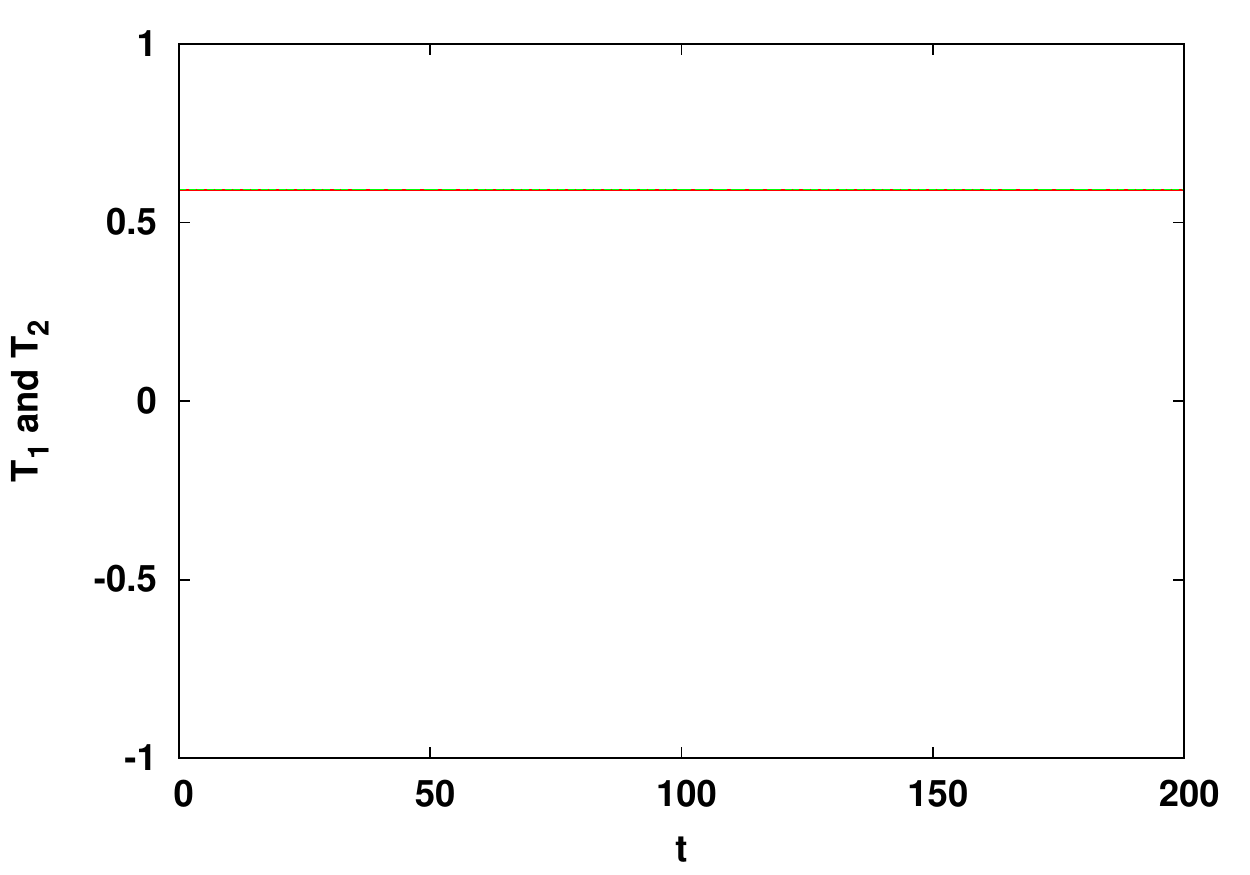}(a)
                		\includegraphics[scale=\SCALE]{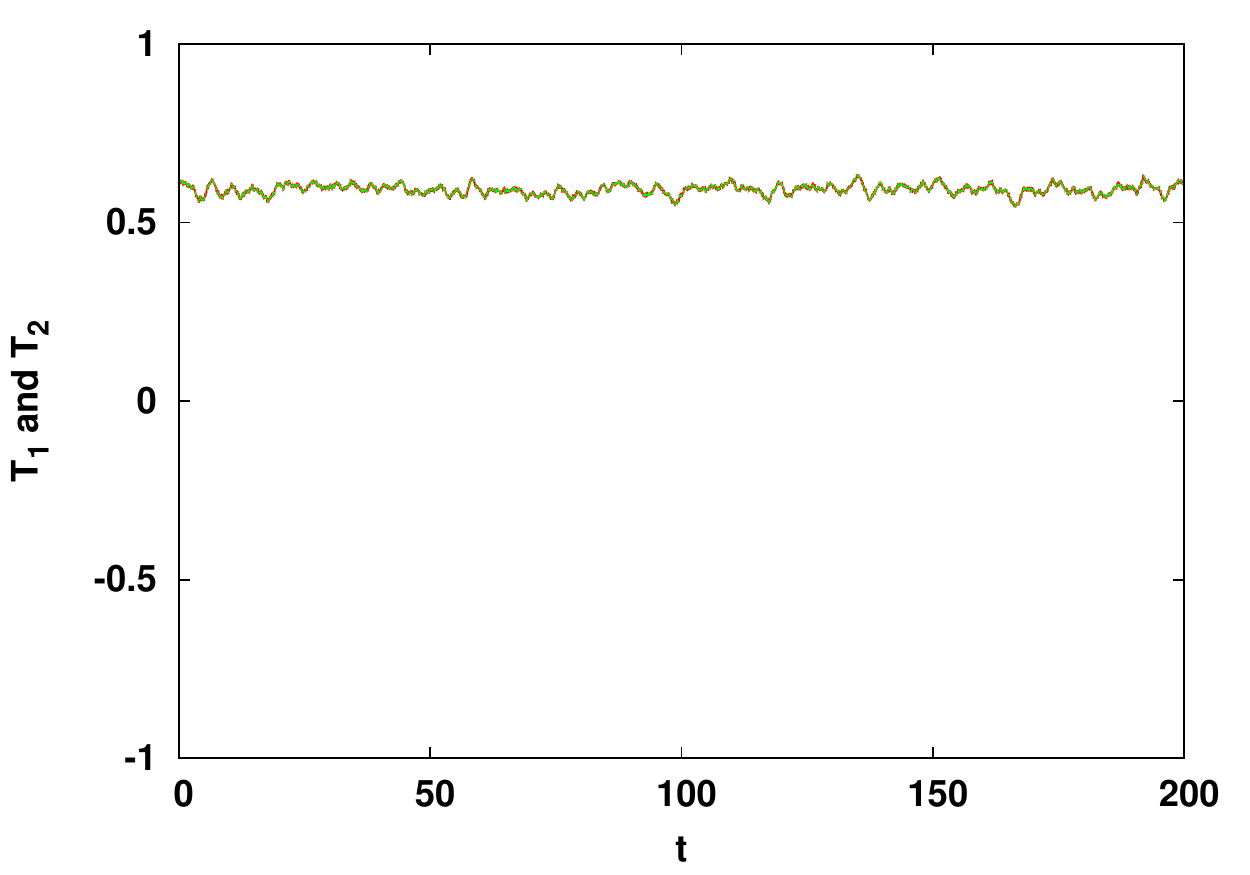}(b)
                		
                		\vspace{1cm}
                		
                		\includegraphics[scale=\SCALE]{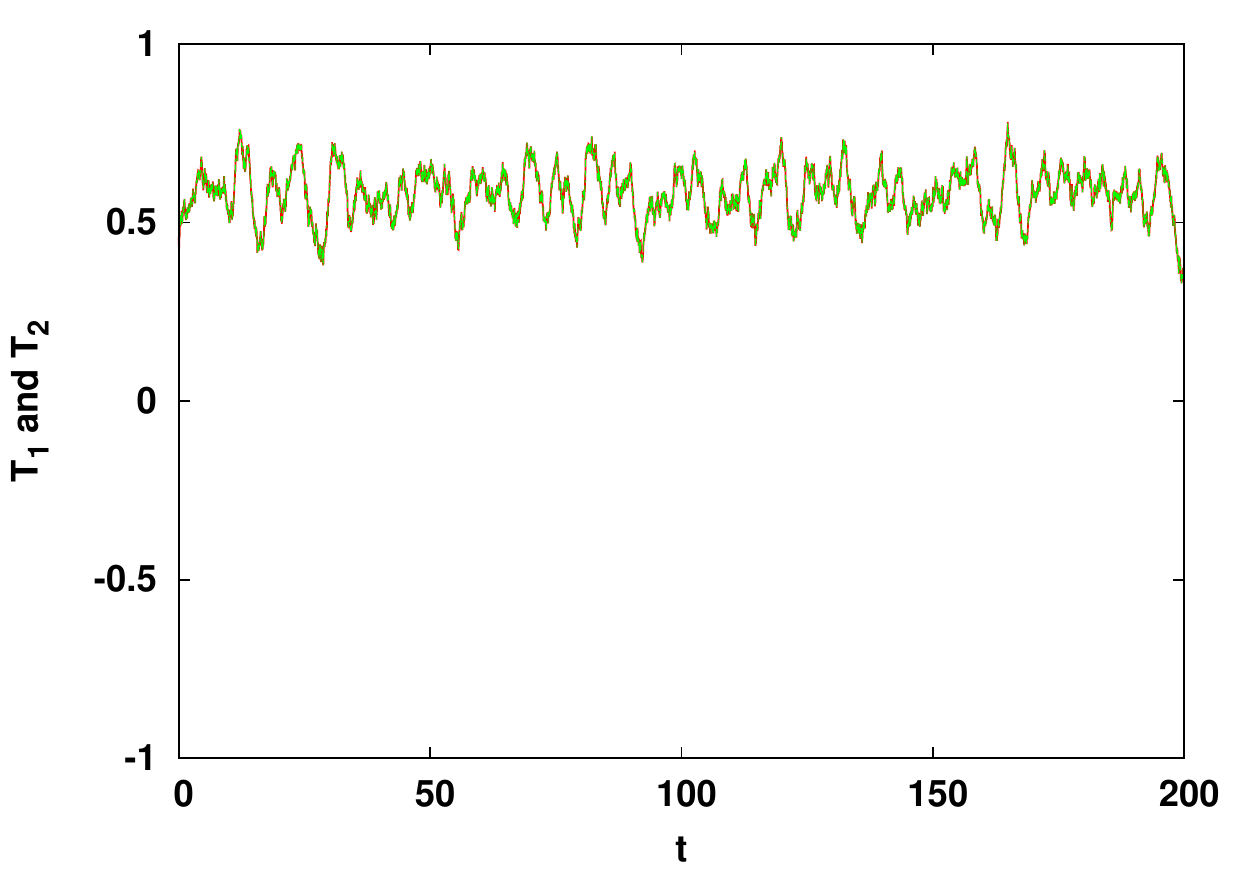}(c)
                		\includegraphics[scale=\SCALE]{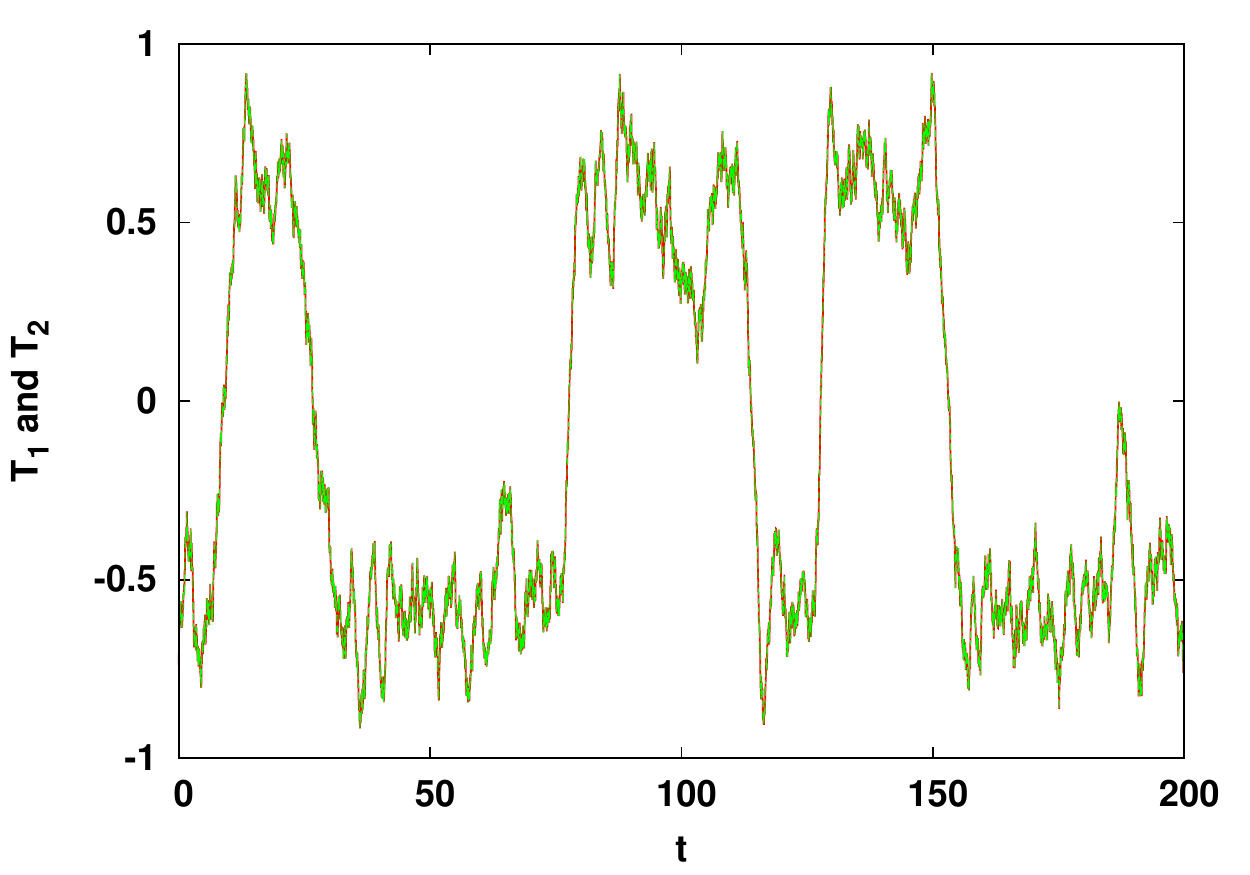}(d)
						\caption{Temporal evolution of $T_1$ and $T_2$, for system parameters $\alpha_{1}=\alpha_{2}=\alpha=0.75, \delta_1=\delta_2=\delta=1, \gamma=0.1$ and initial condition $T_{1}=T_{2}=0.5$. Here the noise strength is (a) $D=0$ (namely without noise), (b) $D=0.01$, (c) $D=0.05$ and (d) $D=0.1$.}\label{a_0p75_ic=p_0p5}
                \end{figure}

                \begin{figure}[H]
                \centering 
                       \includegraphics[scale=\SCALE]{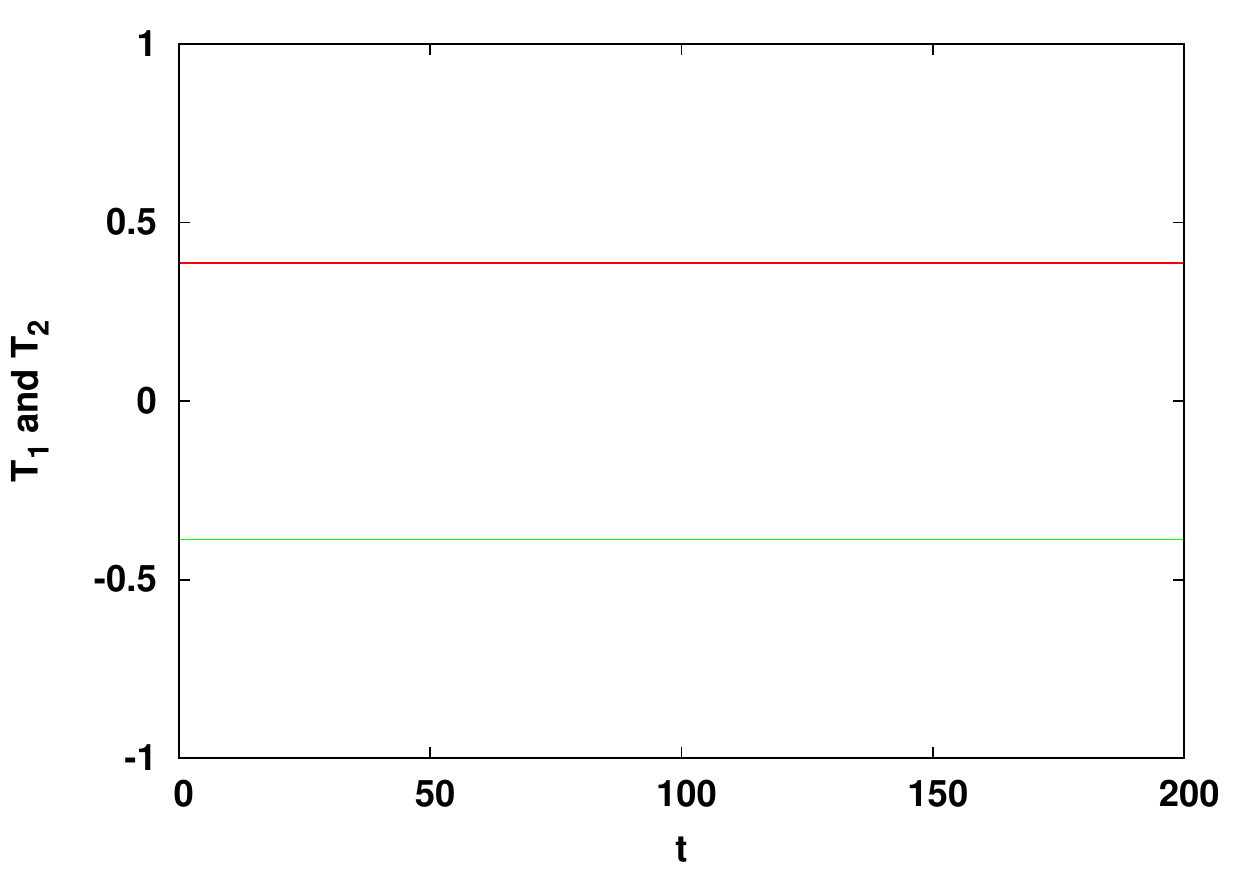}(a)
               		\includegraphics[scale=\SCALE]{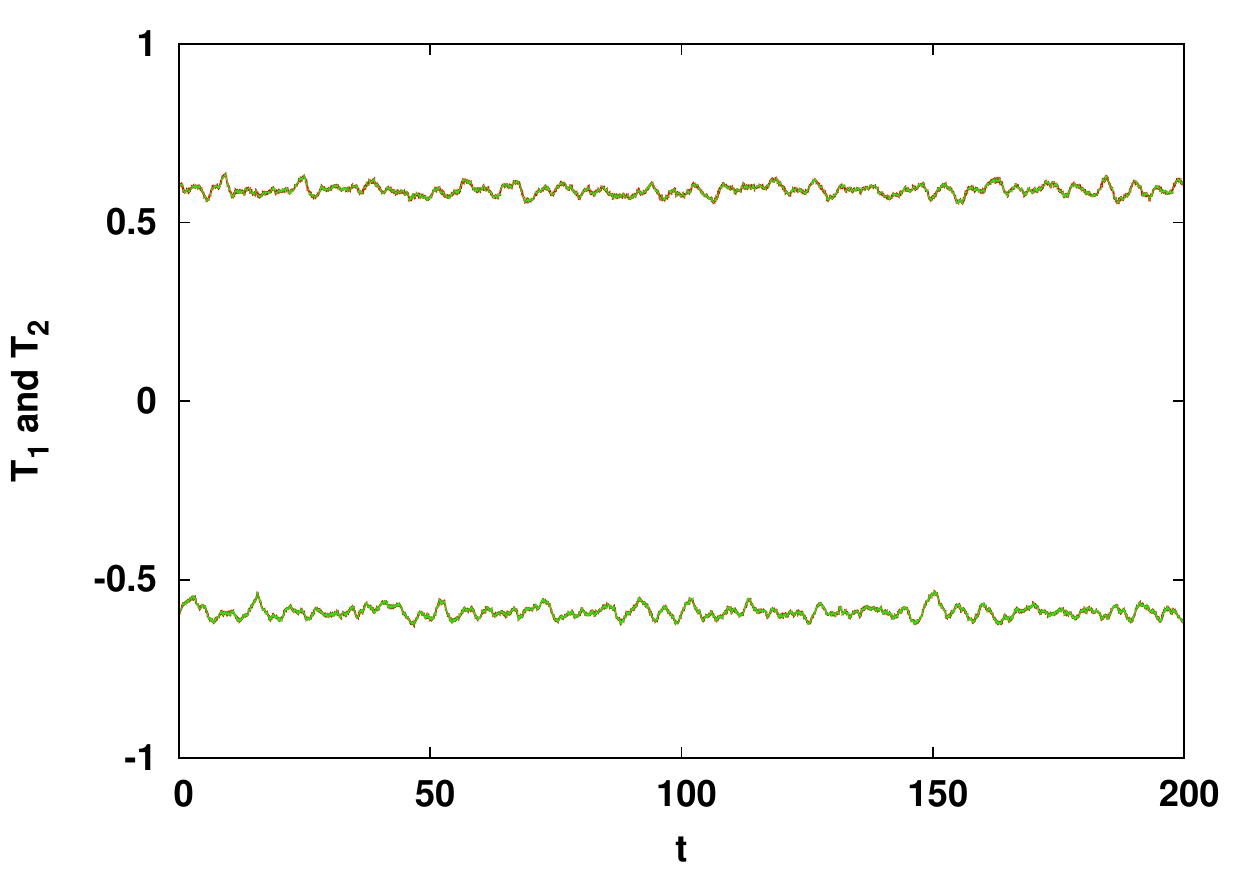}(b)
               		
               		\vspace{1cm}
               		
              		\includegraphics[scale=\SCALE]{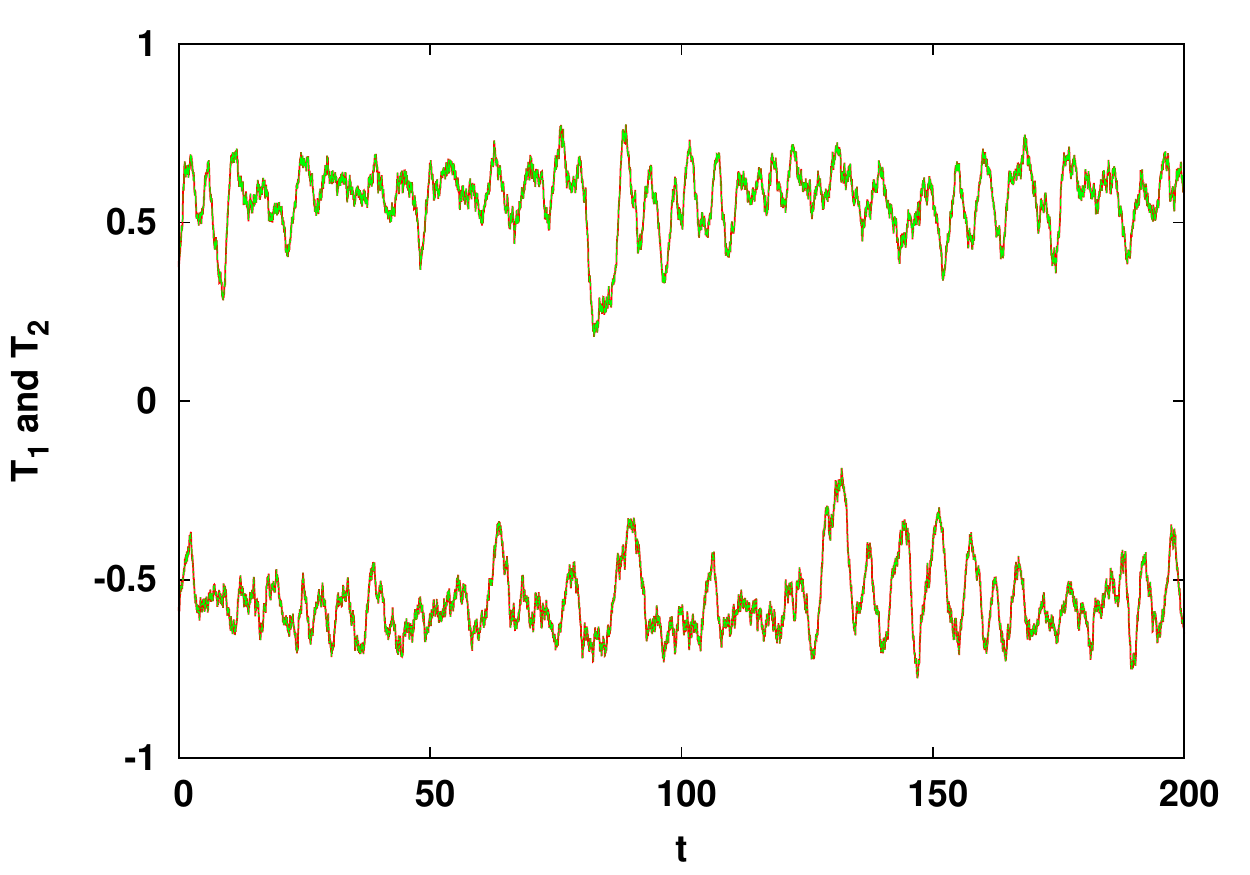}(c)
             			\includegraphics[scale=\SCALE]{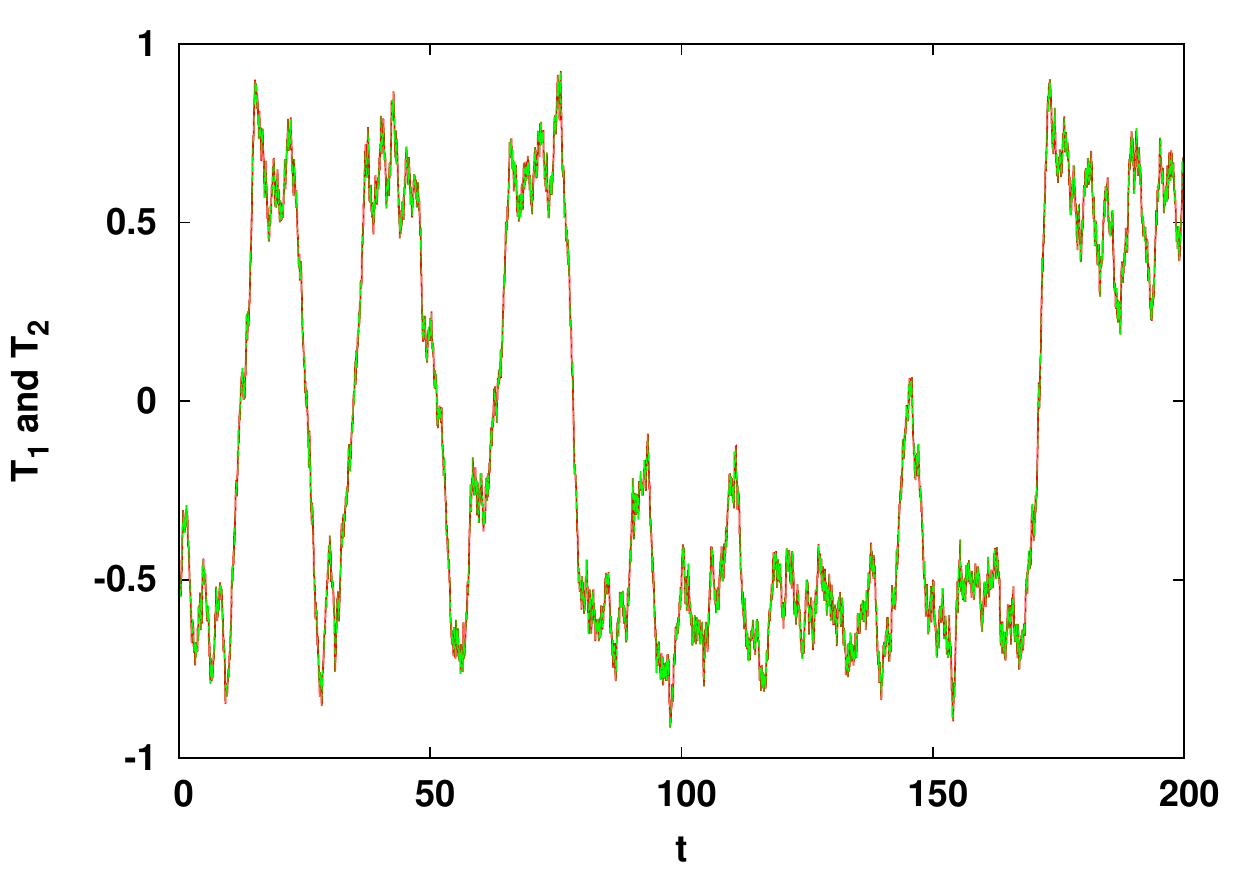}(d)
             			
				\caption{Temporal evolution of $T_1$ and $T_2$, for system parameters $\alpha_{1}=\alpha_{2}=\alpha=0.75, \delta_1=\delta_2=\delta=1, \gamma=0.1$ and initial condition $T_{1}=0.5$, $T_{2}=-0.5$. Here the noise strength is (a) $D=0$ (namely without noise), (b) $D=0.01$, (c) $D=0.05$ and (d) $D=0.1$.}\label{a_0p75_ic=np_0p5}
             \end{figure} 
           
             We now go on to consider another parameter set, $\alpha_{1}=\alpha_{2}=\alpha=0.5, \delta_1=\delta_2=\delta=1, \gamma=0.1$, which yields four steady states: $0.774597$, $-0.774597$, $0.632456$ and $-0.632456$. From initial conditions $T_{1}=T_{2}=0.5$, both sub-systems go to the fixed point at $0.774597$ in the noise-free case (cf. Fig.~\ref{a_0p5_ic=p}a). Under influence of weak and high noise (e.g. $D=0.01, 0.1$), the sub-systems are still attracted to the same state, as evident from Fig.~\ref{a_0p5_ic=p}b-c. When noise strength is very high (e.g. $D=0.2$), there is {\em switching} between $0.774597$ and $-0.774597$ states (cf. Fig.~\ref{a_0p5_ic=p}d). Thus we can infer that these states are more stable compared to the other two states.
			\begin{figure}[H]
             \centering 
                \includegraphics[scale=\SCALE]{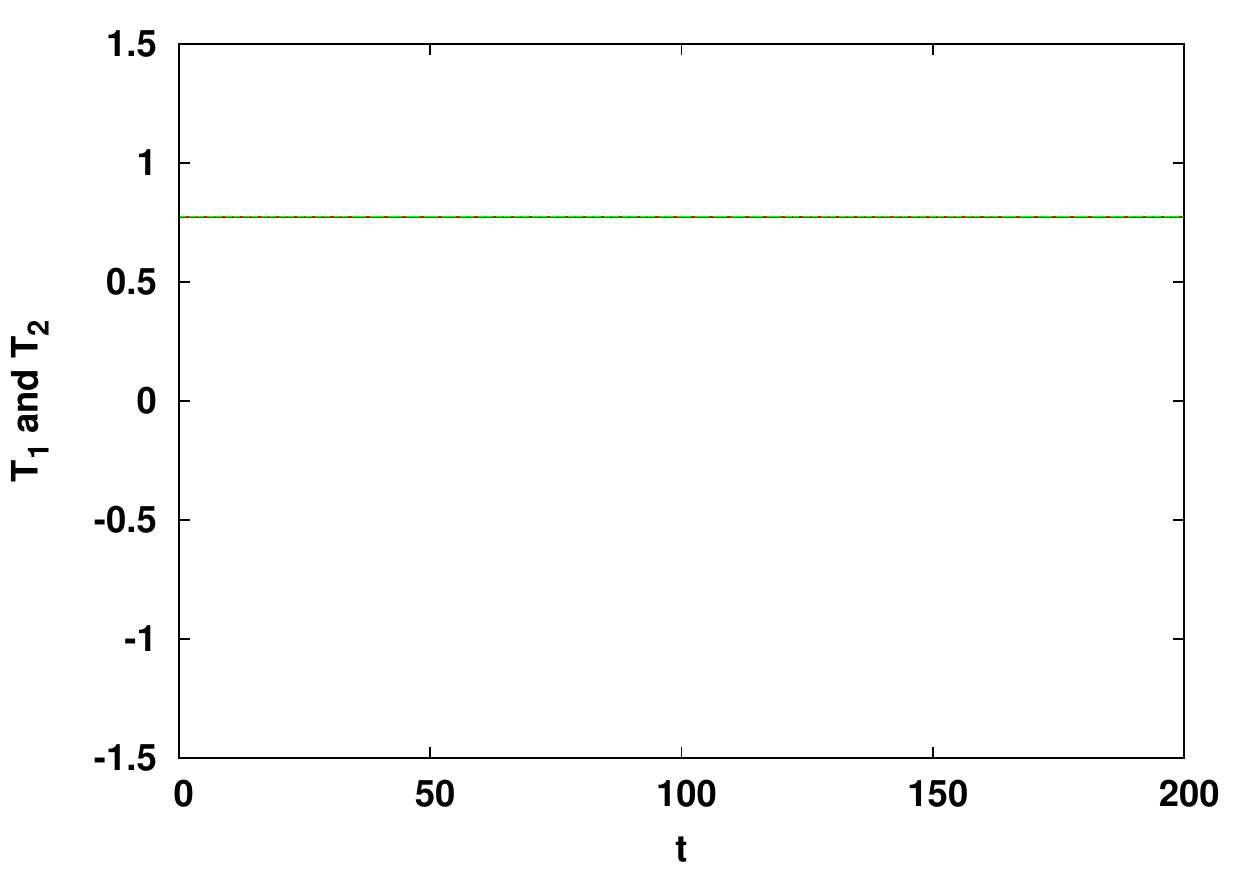}(a)
                \includegraphics[scale=\SCALE]{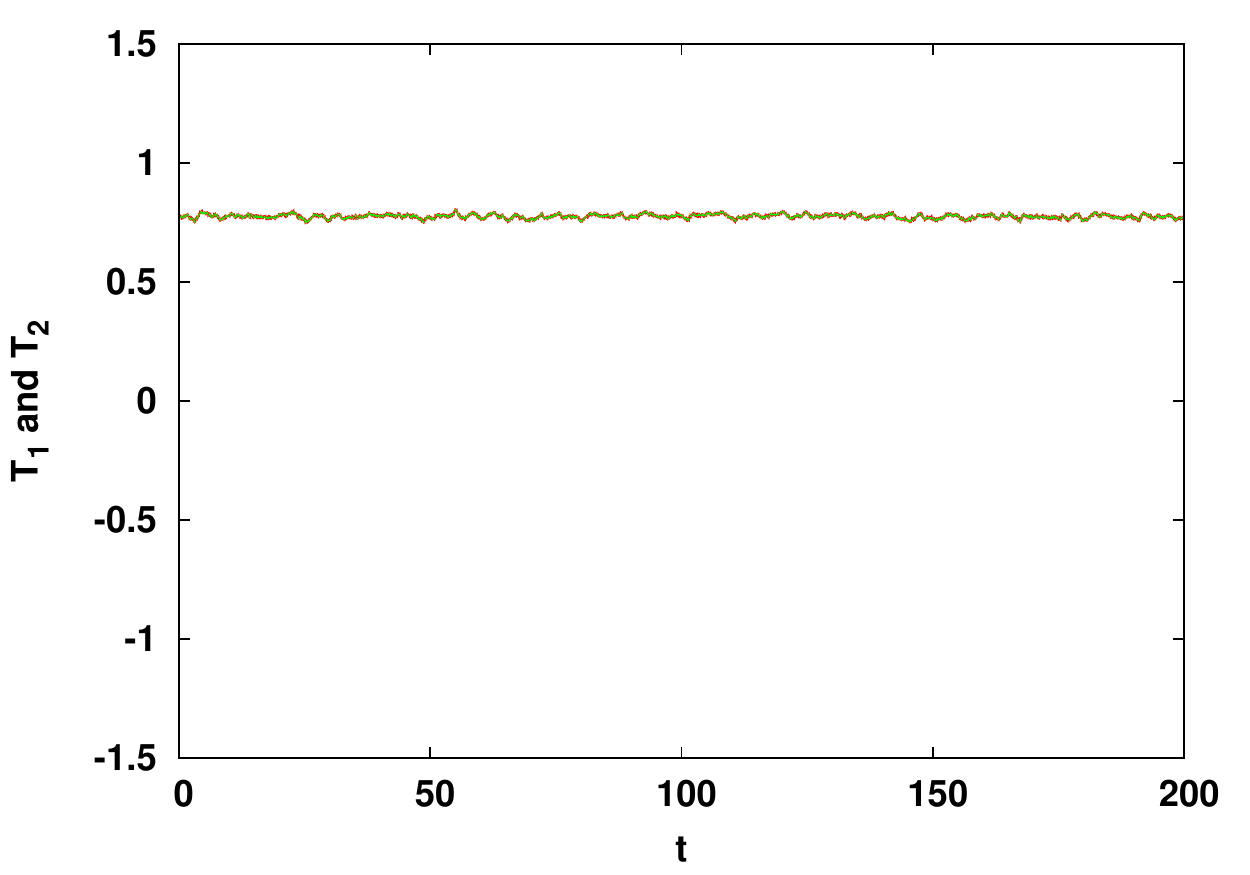}(b)
                
                \vspace{1cm}
                
            	\includegraphics[scale=\SCALE]{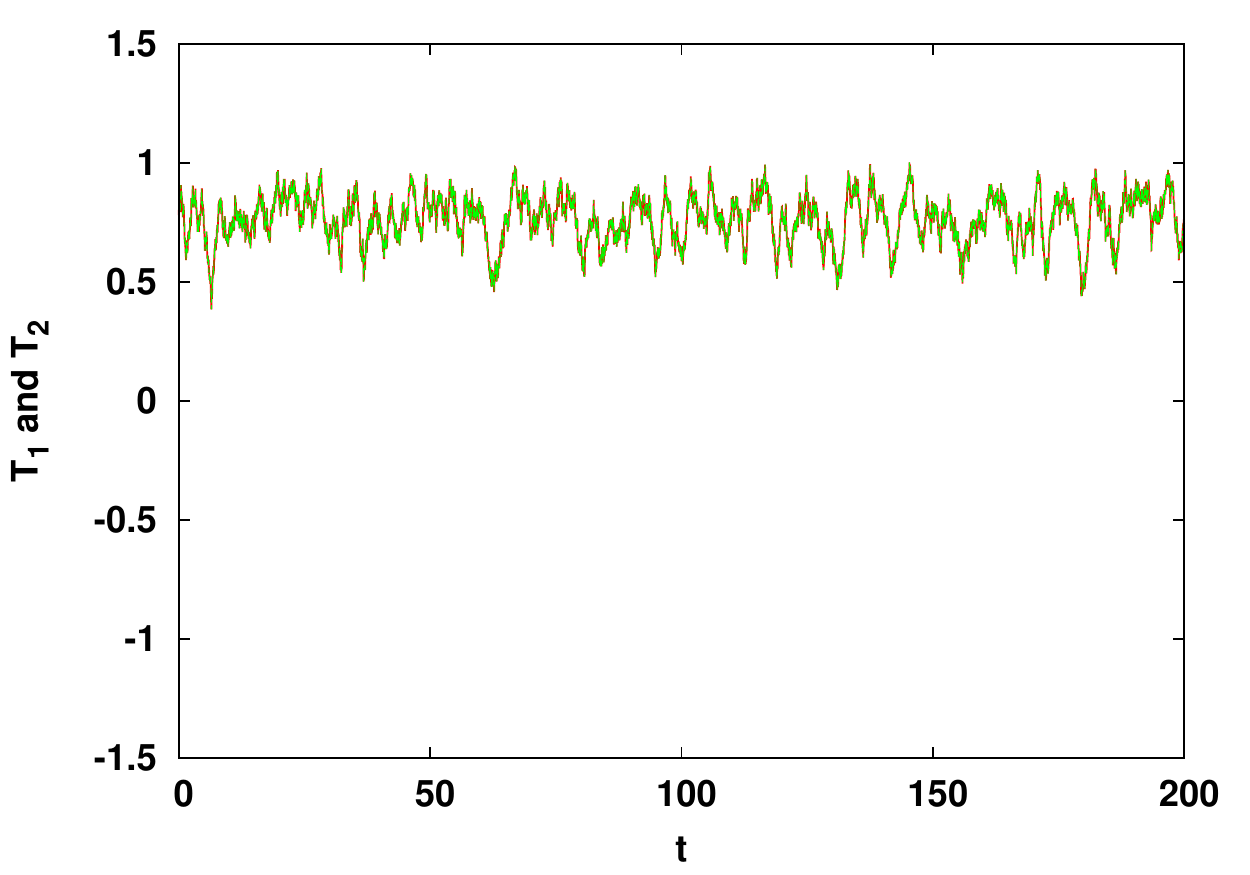}(c)
            	\includegraphics[scale=\SCALE]{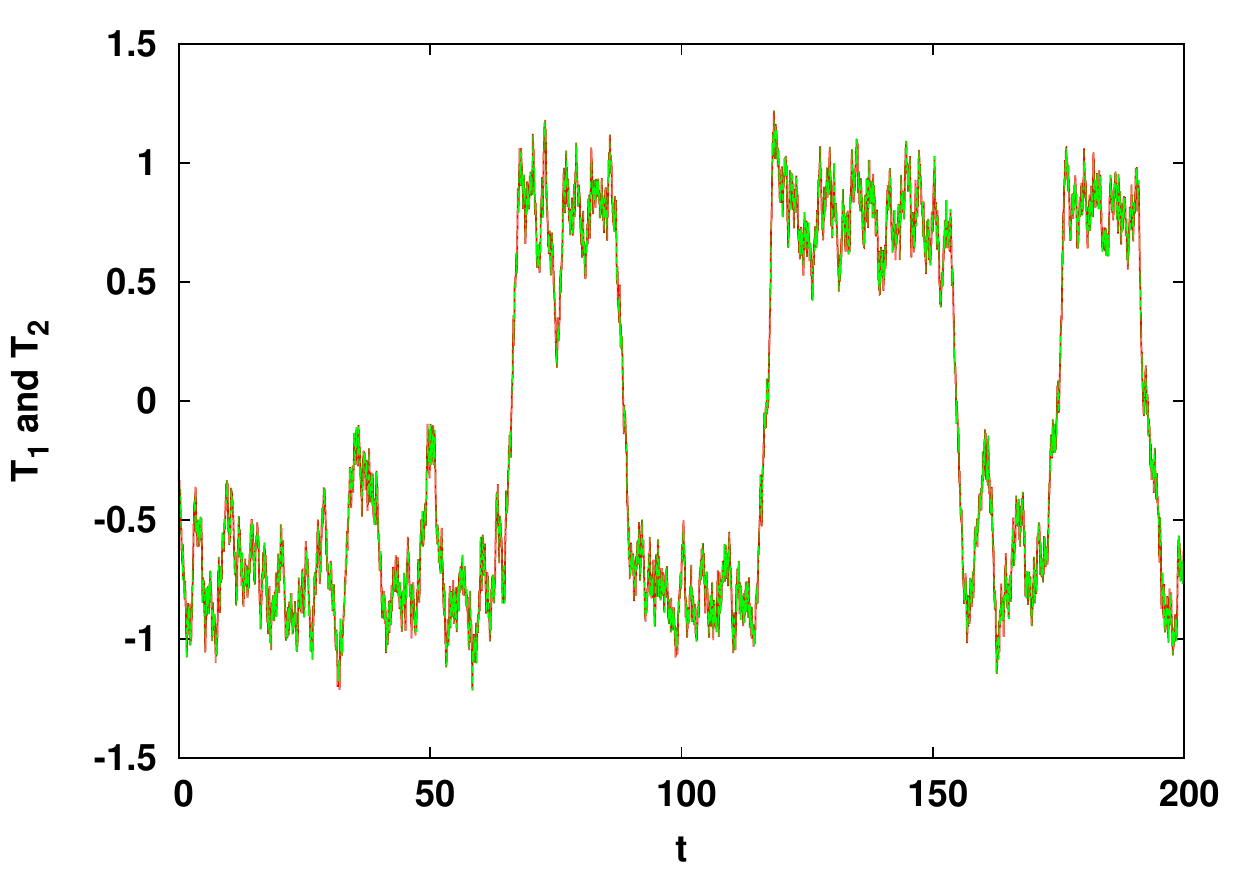}(d)  
            	\caption{Temporal evolution of $T_1$ and $T_2$, for system parameters $\alpha_{1}=\alpha_{2}=\alpha=0.5, \delta_1=\delta_2=\delta=1, \gamma=0.1$ and initial condition $T_{1}=T_{2}=0.5$. Here the noise strength is (a) $D=0$ (namely without noise), (b) $D=0.01$, (c) $D=0.1$ and (d) $D=0.2$.}\label{a_0p5_ic=p}
			
          \end{figure} 

	On examining the initial condition $T_1=1.5$ and $T_2=-1.5$, we observe from Fig.~\ref{a_0p5_ic=np_1p5}(a-b) that when there is no noise or when noise strengths are very weak (e.g. $D=0.01$), each sub-region goes to a different state, namely the sub-regions are attracted to either $0.632456$ or $-0.632456$. For stronger noise (e.g. $D=0.1$) the system evolves to same state, namely both sub-regions evolve to states close to either $0.632456$ or $-0.632456$ (cf. Fig. \ref{a_0p5_ic=np_1p5}c). When noise strength is very high (e.g. $D=0.2$), there is {\em switching} between $0.632456$ and $-0.632456$ states (cf. Fig. \ref{a_0p5_ic=np_1p5}d). Therefore we conclude from Figs. \ref{a_0p5_ic=p}-\ref{a_0p5_ic=np_1p5} that there are four attracting states when the system is under the influence of noise. Note that in \cite{csf_ms} we had seen that lower strengths of self-delay coupling yield more steady states, compared to higher strengths of self-delay coupling. Here too we observe four robust states for $\alpha_1 = \alpha_2 = \alpha =0.5$, and only two robust states for $\alpha_1 = \alpha_2 = \alpha =0.75$. We also observe that for lower values of $\alpha_{1,2}$, larger noise strengths are required to switch between these states.
                
	\begin{figure}[H]
	\centering 
		\includegraphics[scale=\SCALE]{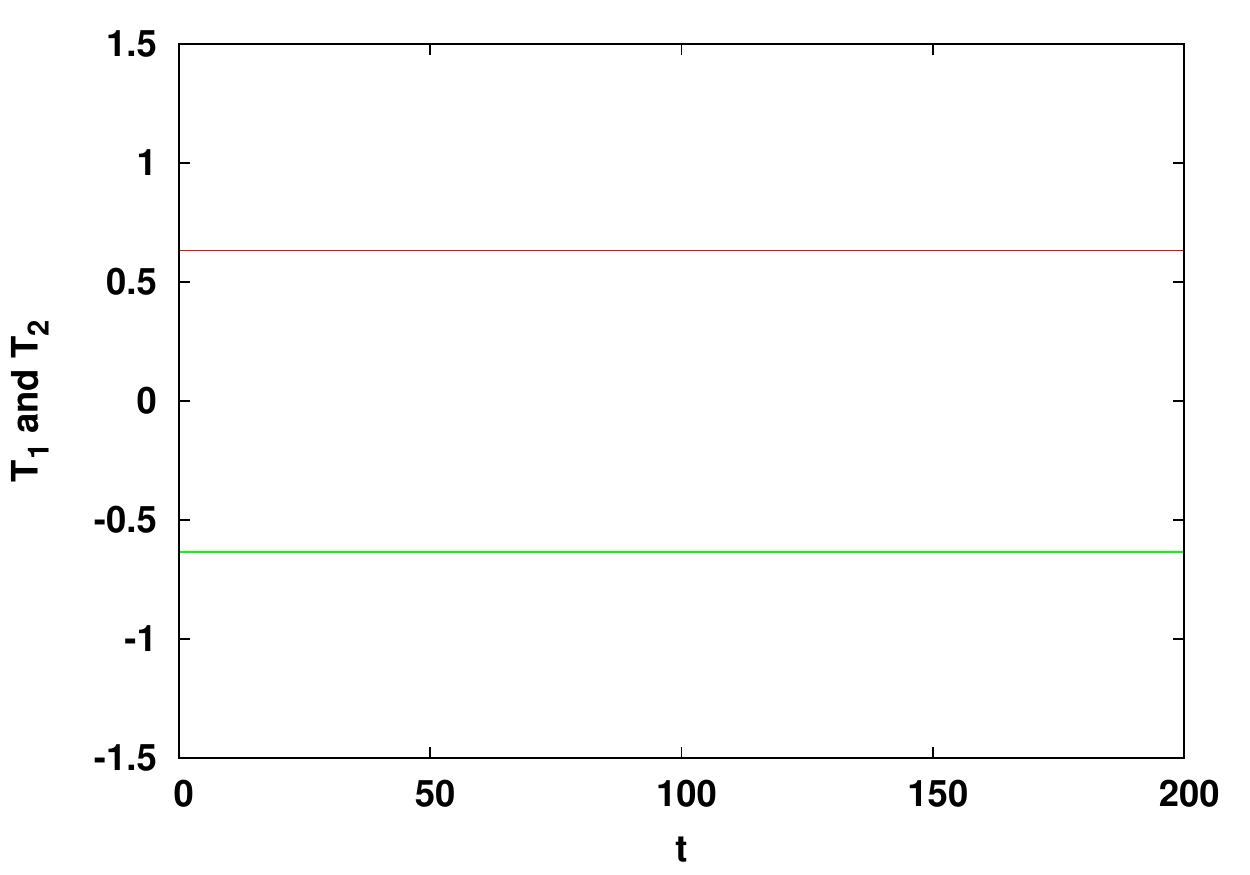}(a)
		\includegraphics[scale=\SCALE]{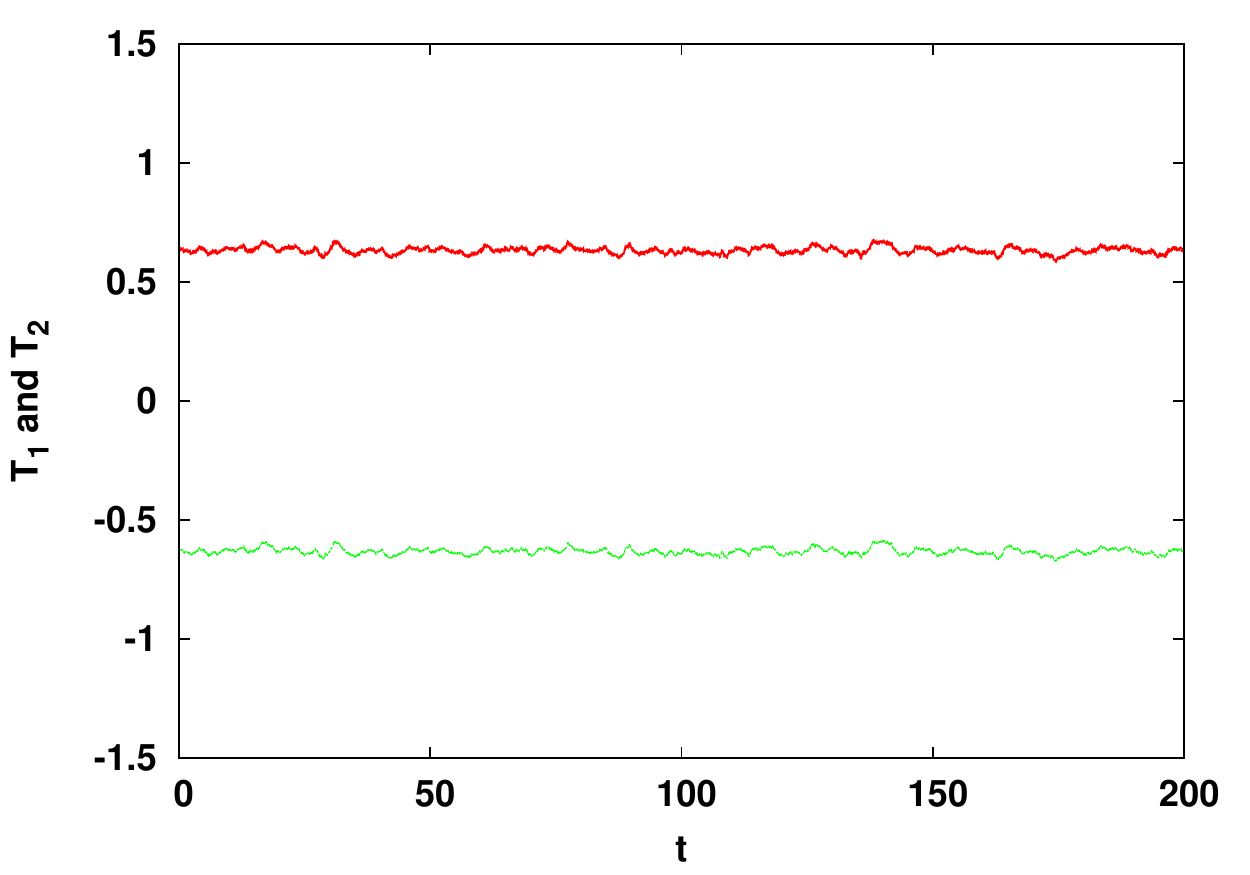}(b)
		
		\vspace{1cm}
		
		\includegraphics[scale=\SCALE]{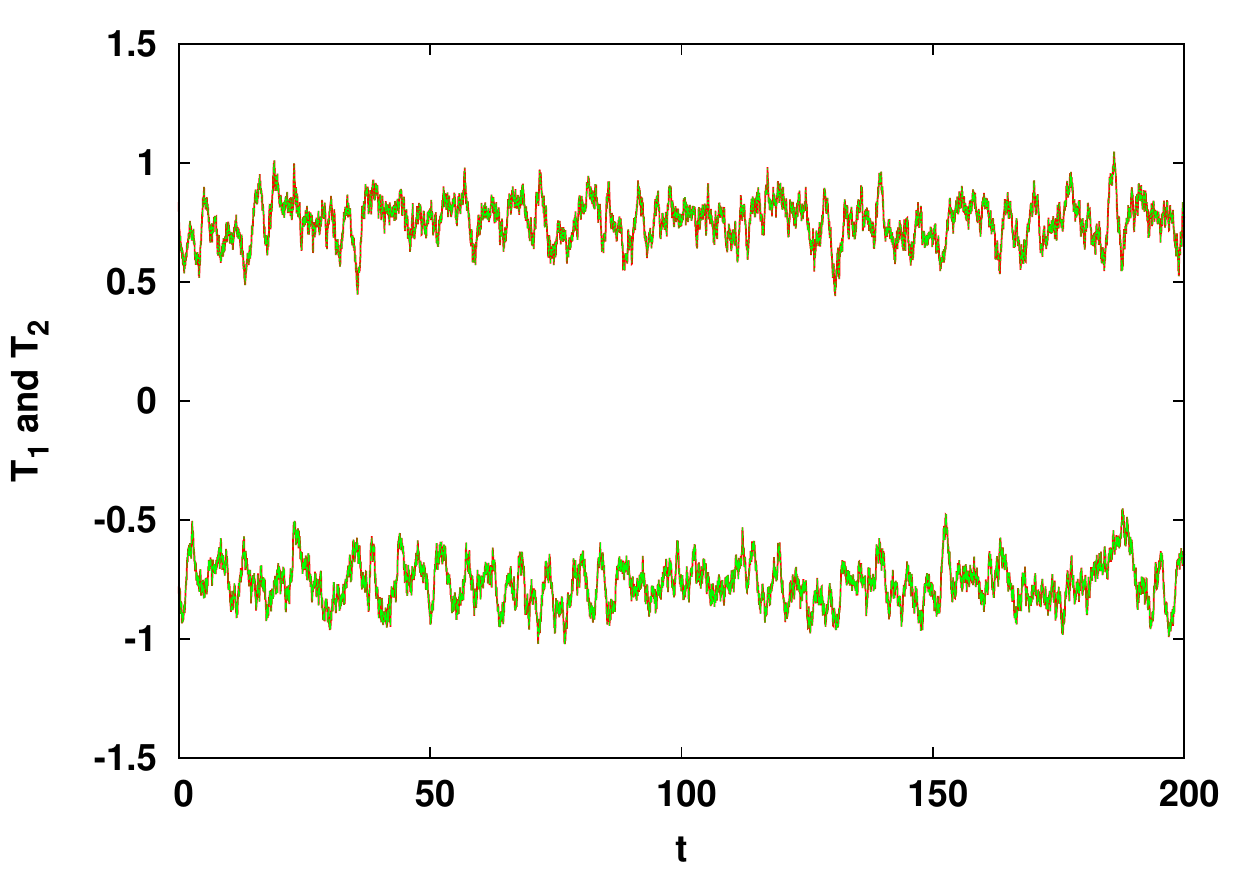}(c)
		\includegraphics[scale=\SCALE]{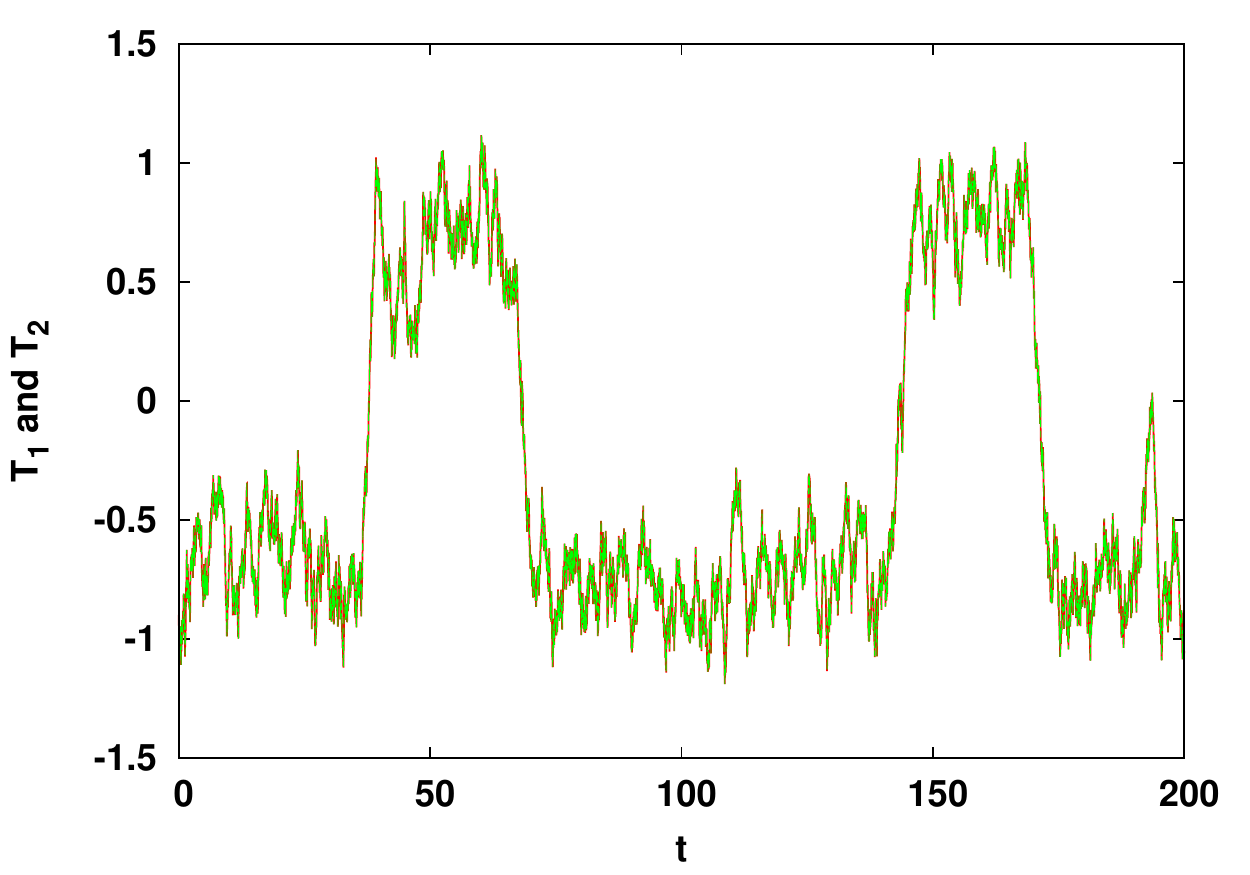}(d)

		\caption{Temporal evolution of $T_1$ and $T_2$, for system parameters $\alpha_{1}=\alpha_{2}=\alpha=0.5, \delta_1=\delta_2=\delta=1, \gamma=0.1$ and initial condition $T_{1}=1.5$, $T_{2}=-1.5$. Here the noise strength is (a) $D=0$ (namely without noise), (b) $D=0.01$, (c) $D=0.1$ and (d) $D=0.2$.}\label{a_0p5_ic=np_1p5}
		
	\end{figure} 
	
   \section{Discussions}
  
  	%In Ref.\cite{csf_ms}
	  	 We considered a system of coupled delayed action oscillators modelling the El Ni\~{n}o effect, and studied the dynamics of the sea surface temperature (SST) anomaly. The existence and stability of the solutions arising in this model depend on three parameters: self delay, delay and inter-region coupling strengths. In our work we explore the dynamics in the space of these parameters. The emergence or suppression of oscillations in our models is a dynamical feature of utmost relevance, as it signals the presence or absence of ENSO-like oscillations. Note that in contrast to the well-known low order model of ENSO, the recharge oscillator \cite{recharge2} and its important stochastic extensions \cite{bianucci}, where the influence of the neighbouring regions on the region of interest is modelled as external noise, we consider neighbouring regions as a coupled deterministic dynamical systems. Different parameters yield a rich variety of dynamical patterns in our model, ranging from steady states and homogeneous oscillations to irregular oscillations, without explicit inclusion of noise.
	  	 
	  	  In our earlier results \cite{csf_ms} for identical sub-regions, we had typically observed a co-existence of amplitude and oscillator death behavior for low delays, and heterogeneous oscillations for high delays, when inter-region coupling is weak. For moderate inter-region coupling strengths one obtained homogeneous oscillations for sufficiently large delays and amplitude death for small delays. When the inter-region coupling strength was large, oscillations were suppressed altogether, implying that strongly coupled sub-regions do not yield ENSO-like oscillations. Further we observed that larger strengths of self-delay coupling favoured oscillations, while oscillations died out when the delayed coupling was weak. This indicated again that delayed feedback, incorporating oceanic wave transit effects, was the principal cause of oscillatory behaviour. So the effect of trapped ocean waves propagating in a basin with closed boundaries is crucial for the emergence of ENSO-like oscillations. The non-uniformity in delays, and difference in the strengths of the self-delay coupling of the sub-regions, was also investigated. As in the uniform case, larger delays and self-delay coupling strengths lead to oscillations, while strong inter-region coupling killed oscillatory behaviour. The difference between the uniform case and the non-uniform system, was that amplitude death and homogeneous oscillations are predominant in the former, while oscillator death and heterogeneous oscillations were commonly found in the latter. Interestingly, we also found that when one sub-region had low delay and another had high delay, under weak coupling the oscillatory sub-region induced oscillations in sub-region that would have gone to a steady state if uncoupled.
	  	  
	  	   Moreover, we have also explored the robustness of the different dynamical states under noisy evolution, in order to gauge which set of attractors are typically expected to arise when the system evolves under the influence of external perturbations. Typically we find that the noisy system evolves to a sub-set of the attractors found in the deterministic system, and those attractors can be considered robust under noise. Often when noise is very weak, the system is attracted to states close to the noise-free case. However when noise is stronger, the system switches randomly between the attractors. Using this method of gauging the robustness of the different attractors in our multi-stable system, we find that lower strength of self-delay coupling yields a larger number of robust states, than stronger self-delay coupling. Further, larger noise strengths are required to switch between these states, when the strength of self-delay coupling is low.
	  	  
	  	  We then investigated the basins of attraction of the different dynamical attractors arising in our model. Typically, the number of distinct attractors and their basins of attraction depend upon the values of  parameters. For instance, when $\alpha_{1}=\alpha_{2}=0.75, \delta=1$, we find four steady states for $\gamma=0.1$ and two steady states for $\gamma\ge0.2$. The value of the fixed points depend on the values of the inter-region coupling strength $\gamma$. For the typical case of $\alpha_1 \ne \alpha_2$, each sub-region has two fixed points and two oscillator states, with the attractors being different in the two regions. Further, generically, in such cases there is a complex co-existence of attractors. 
	  
	  	  Now, several agencies such as the National Oceanic and Atmospheric Administration (NOAA) in the United States, monitor the sea surface temperatures in various regions, five degrees of latitude on either side of the equator, with  Ni\~{n}o 1-2 region located in the band 80W--90W, Ni\~{n}o 3 region in 90W--150W, and Ni\~{n}o 4 region in 160E--150W. The Ni\~{n}o 3.4 region (120W--170W) is often the primary focus for monitoring and predicting El Ni\~{n}o. When the three-month SST average for the area is more than 0.5${\degree}$C above (or below) normal for that period, then an El Ni\~{n}o (or La Ni\~{n}a) is considered to be in progress.
	  	  	
	  	  	How our model can explain the $0.5\degree$C 
	  	  	criterion used for the forecasting, we show by rescaling our result and comparing it with observations. We consider two regions along the equator, where the first region extends from $90\degree$ West to $150\degree$ West
	  	  	(Ni\~{n}o 3 region) with the mid-point being $120\degree$ West and the second region extends from $150\degree$ West to $160\degree$ East (Ni\~{n}o 4 region) with the mid-point being $175\degree$ West. The western Pacific boundary is at $120\degree$ East. This gives angular separation of $120\degree$ and $65\degree$ of longitude for the waves to travel, for the two regions respectively, and corresponds to a distance $120(2\pi/360)\times r_{Earth} = 13.35\times10^6m$ and $65(2\pi/360)\times r_{Earth}=7.23\times10^6m$ for the two regions, where $r_{Earth}=6.37\times10^6 m$. Speed of the Kelvin wave is $1.4$ms$^{-1}$ and $0.47$ms$^{-1}$ for Rossby wave \cite{AmJPhys}. These values of speed gives $13.35\times10^6m/0.47ms^{-1}=329$ days for the Rossby propagation to
	  	  	the western boundary, and a further $13.35\times10^6m/1.4ms^{-1}=110$ days for the return of the Kelvin waves, thus total delay of transient time 
	  	  	$\Delta$=$439$ days for the first region.
	  	  	For the second region it gives  $7.23\times10^6m/0.47ms^{-1}=178$ days for the Rossby propagation to the western boundary, and a further $7.23\times10^6m/1.4ms^{-1}=59$ days for the return of the Kelvin waves, thus total delay of transient time $\Delta$= $237$ days. 
	  	  
	  	  	In the Eq.~2 the coupling constant and temperature are introduced with the following scaling: $k=\delta/\Delta$  and $T= T^{'}\sqrt{k/b}$, and $b$ is given by $b=k(T^{'}/T)^2$. For the first region ( Ni\~{n}o 3 region) maximum anomaly temperature ($T_1$) is on average $2.11\degree$C and for the second region ( Ni\~{n}o 4 region) maximum anomaly temperature ($T_2$) is on average $1.15\degree$C (these values we have 
	  	  	received from the data produced by NOAA \cite{data3,data4}).
	  	  	Parameter set $\alpha_{1}=0.75$, $\alpha_{2}=0.5$, $\gamma=0.1$ and
	  	  	$\delta=2$, allow us to calculate $k_1=1.64/$years, $k_2=3.03/$years,
	  	  	$b_1=0.43\degree$C$^{-2}/$years and $b_2=1.66\degree$C$^{-2}/$years. 
	  	  	So, now if we consider the $0.5\degree$C criterion used for the forecasting as $T_1=0.5\degree$C and $T_2=0.5\degree$C, then after rescaling, equivalents to the dimensionless temperatures in two regions in the model are $T^{'}_1=0.256$ and $T^{'}_2=0.37$. Thus, we can find the 
	  	  	the criterion location on the diagram of basins of attraction in Fig. \ref{fp_os_a1_p65}(b)(left). It belongs to the green basin which corresponds to the El-Nino state, in the vicinity of the boundary with the gray basin associated with oscillation between El-Nino and La-Nino
	  	  	states. Hence, the model can reproduce the $0.5\degree$C criterion revealed from observations.
	  	  	
	  	  	Additionally, our modelling result suggests that instead of the single value criterion (as $0.5\degree$C),
	  	  	an interval should be used as criterion to estimate the El-Nino or La-Nino progress.  According to the model result,
	  	      if temperature anomaly of $T_1$ is in the range $0.5\degree$C$<T_1<1.6\degree$C (corresponding $0.32<T^{'}_1<0.84$) and $T_2$ is in the range $-0.06\degree$C$<T_2<2\degree$C (corresponding $-0.04<T^{'}_2<1.52$),then  El Ni\~{n}o is considered to be in progress.
	  	  	If temperature anomaly $T_1$ is in the range $-1.6\degree$C$<T_1< -0.5\degree$C (corresponding $-0.84<T^{'}_1<-0.32$) and $T_2$ is in the range $-2\degree$C$<T_2<0.06\degree$C (corresponding $-1.52<T^{'}_2<0.04$), then La Ni\~{n}o is considered to be in progress. In other range of temperature anomalies, ENSO(successive El Ni\~{n}o and La Ni\~{n}o) episodes are considered to be in progress.
	  	  	
	  	  	Hence, the basins of attraction for the different steady states and oscillatory states in our model may help in understanding patterns in the sea surface temperatures anomalies in monitored coupled sub-regions. Further, our mapping of the basins of attraction might be helpful for forecasting of El Ni\~{n}o (or La Ni\~{n}a) progress, as it indicates the combination of initial SST anomalies in the sub-regions that can result in a El Ni\~{n}o/La Ni\~{n}a episodes.
	  	  	
	  	   In summary then, we have explored a simple model based on coupled delayed action oscillators modelling the ENSO-like oscillations, and studied the dynamical patterns of the sea surface temperature (SST) anomaly. Specifically we have presented the existence, stability and basins of attraction of the solutions arising in the model system, for different representative parameter sets. Thus our dynamical model may help provide a potential framework in which to understand patterns in the SST anomalies in different coupled sub-regions, which is an important feature that has not yet been sufficiently explored.
  
  \bigskip

  	\noindent
  	{\bf Acknowledgements}
  	
  	CM would like to acknowledge the financial support from DST INSPIRE Fellowship, India.

  %\bigskip

  	\end{document}